\documentclass{aa}

\usepackage{graphicx}
\usepackage{txfonts}
\usepackage{lastpage}

\usepackage{hyperref}
\hypersetup{
    colorlinks,
    linkcolor={blue},
    citecolor={blue},
    urlcolor={blue}
}
\usepackage{multicol}
\usepackage{multirow}

\usepackage{subcaption}

\usepackage{threeparttable}
\usepackage{orcidlink}

\graphicspath{{Figures/}}
\newlength{\fullwidth}
\newlength{\halfwidth}
\begin{document}

   \title{Simulating the LOcal Web (SLOW) \\V. Thermodynamic Properties and Evolution of Local Galaxy Clusters}

   \author{Elena Hernández-Martínez\orcidlink{0000-0002-1329-9246}\thanks{\email{hernandez@usm.lmu.de}}\fnmsep
          \inst{1}
          \and
          Klaus Dolag\inst{1,2}
          \and
          Ulrich P. Steinwandel\orcidlink{0000-0001-8867-5026}\inst{2}
          \and
          Jenny G. Sorce\inst{3,4}
          \and
          Théo Lebeau\inst{4}
          \and
          Nabila Aghanim\inst{4}
          \and
          Benjamin Seidel\inst{1}
         }

   \institute{Universit\"ats-Sternwarte, Fakult\"at f\"ur Physik, Ludwig-Maximilians-Universit\"at M\"unchen, Scheinerstr.1, 81679 M\"unchen, Germany
         \and
             Max Planck Institute for Astrophysics, Karl-Schwarzschild-Str. 1, D-85741 Garching, Germany
                      \and
   Univ. Lille, CNRS, Centrale Lille, UMR 9189 CRIStAL, F-59000 Lille, France
         \and
   Universit\'e Paris-Saclay, CNRS, Institut d'Astrophysique Spatiale, 91405, Orsay, France   
 }
\authorrunning{Elena Hernández-Martínez et al.}  
\titlerunning{Simulating the LOcal Web V}  

   \date{Accepted XXX. Received YYY; in original form ZZZ}

 
  \abstract
{Galaxy clusters are the largest gravitationally bound structures in the Universe, serving as key laboratories for studying structure formation and evolution. The intracluster medium (ICM), composed of hot plasma, dominates their baryonic content and is primarily observable in X-rays. The thermodynamic properties of the ICM, including pressure, temperature, entropy, and electron density, provide crucial insights into the physical processes shaping these systems, from accretion and mergers to radiative cooling and feedback.}
{We investigate the thermodynamic properties of galaxy clusters within the Simulating the LOcal Web (SLOW) constrained simulations, which reproduce the observed large-scale structure of the local Universe. Our goal is to assess the reliability of these simulations in reproducing the observed ICM profiles of individual clusters and to explore the connection between cluster formation history and core classification.}
{We extract three-dimensional thermodynamic profiles from the simulations, assuming spherical symmetry. These profiles are directly compared to deprojected X-ray and Sunyaev–Zel'dovich (SZ) observational data for a sample of local clusters, including systems classified as solid cool-core (SCC), weakly cool-core (WCC), and non-cool-core (NCC) systems. Additionally, we analyze the mass assembly history of the simulated cluster counterparts to establish links between their formation pathways and present-day ICM properties. } 
{The simulations successfully reproduce the global thermodynamic profiles of observed clusters, particularly in the pressure and temperature distributions within \( r_{500} \). The electron density and entropy profiles of cool-core (CC) clusters show some discrepancies, likely due to resolution limitations and the treatment of feedback processes. We find that CC clusters typically assemble their mass earlier, while NCC clusters experience more extended merger-driven growth. WCC clusters exhibit intermediate accretion histories, suggesting an evolutionary transition between CC and NCC states.  }
{Our results demonstrate that constrained simulations provide a powerful tool for linking cluster formation history to present-day ICM properties. While the large-scale structure and bulk thermodynamic profiles are well reproduced, further refinements in subgrid physics as well as higher resolution are needed to improve the agreement in cluster core regions. These findings offer new insights into the evolution of cluster cores and their observational classification, reinforcing the importance of constrained simulations for studying galaxy clusters in a cosmological context.  }


   \maketitle
%




\section{Introduction}
\label{Introduction}
Galaxy clusters, the largest virialized and gravitationally bound structures in the Universe, are vital laboratories for studying cosmic structure formation and evolution  \citep[e.g.,][]{2011ARA&A..49..409A,2012ARA&A..50..353K,2013A&A...550A.131P,2014A&A...571A..20P,2024arXiv241010942H}. A key component of galaxy clusters is the intracluster medium (ICM), a hot, weakly collisional diffuse plasma that dominates their baryonic mass \citep[e.g.,][]{Carilli2001}. With temperatures reaching tens of millions of kelvin, the ICM emits X-rays primarily through thermal bremsstrahlung radiation, making it observable with modern X-ray telescopes. The thermodynamic properties of the ICM, including temperature, electron density, entropy, and pressure, provide critical insights into the physical processes governing cluster formation and their applications as cosmological probes \citep[e.g.,][]{2006ApJ...640..691V,2013A&A...551A..23E, 2019SSRv..215...25P}. Understanding these properties is essential for linking observations to theoretical models of large-scale structure growth.

The ICM’s thermodynamic profiles are shaped by the interplay of gravitational processes, such as accretion and mergers, and non-gravitational mechanisms, including radiative cooling and feedback from stars and active galactic nuclei (AGN). For instance, entropy profiles reveal the balance between gravitational heating and energy input from non-gravitational processes. Pressure profiles are key to connecting the thermal energy of the gas to the cluster’s mass \citep[e.g.,][]{2005RvMP...77..207V, 2010A&A...517A..92A}. Despite significant progress, many aspects of ICM physics remain unclear, such as the precise role of AGN feedback in regulating cluster cores and the mechanisms responsible for the observed deviations from self-similar scaling, particularly in disturbed systems \citep[e.g.,][]{2007ARA&A..45..117M, 2014MNRAS.438..195P, 2014MNRAS.441.1270L}. 

In our current understanding of galaxy cluster formation, the dynamical state of a cluster significantly impacts its ICM profiles. Relaxed clusters, typically characterized by symmetric profiles and cool cores, exhibit stable thermodynamic properties where cooling and heating are in balance. Conversely, disturbed clusters—often believed to be the result of recent mergers—show irregular temperature and density distributions, reflecting their turbulent formation histories \citep[][]{mccarthy2008towards, rossetti2011, rasia2015cool, hahn2017}. The diversity in the observed cluster profiles highlights the importance of understanding the relationship between the dynamical state of a cluster and the physical processes that shape its ICM \citep[e.g.,][]{2015ApJ...813L..17R,2017MNRAS.471.1088B}. These complexities pose challenges for both observational studies and simulations to understand and replicate the behavior of the cluster under a wide range of conditions.

While statistical studies of galaxy cluster populations provide valuable insights into general trends, they often obscure the unique features of individual systems. Constrained simulations offer a complementary approach, enabling the study of specific clusters \citep[e.g.][]{1987ApJ...323L.103B, 2005MNRAS.363...29D,2013MNRAS.432..894J, 2016MNRAS.460.2015S} by reconstructing their initial conditions from observational data, such as galaxy density fields \citep[e.g.,][]{1991ApJ...380L...5H, 2012MNRAS.427L..35K, 2013MNRAS.435.2065H, 2016MNRAS.455.3169L} or peculiar velocity measurements \citep[e.g.,][]{2010MNRAS.406.1007L, 2013MNRAS.430..888D, 2013AJ....146...69C, 2016MNRAS.455.2078S}. This approach allows for one-to-one comparisons between simulated and observed clusters, providing a powerful tool to investigate how individual formation histories influence ICM properties. Constrained simulations thus offer a unique perspective on the diversity of cluster profiles, shedding light on the physical processes driving deviations from expected behaviors.

In this study, we investigate the thermodynamic properties, temperature, electron density, entropy, and pressure, of the ICM of galaxy clusters within the local Universe simulation SLOW \citep{2023A&A...677A.169D}. Using constrained simulations, we analyze these properties on a cluster-by-cluster basis and compare them with observational data at low redshift. We expand on previous work by \citet{2024A&A...687A.253H}, which showed that the SLOW constrained simulations accurately reproduce the integrated properties of galaxy clusters. Subsequent studies have shown that these simulations are useful for exploring CMB large-scale anomalies \citep[][]{2024A&A...692A.180J}, cosmic rays in local galaxy clusters \citep[][]{2024A&A...692A.232B}, dark matter candidates \citep[][]{2024A&A...691A..38S}, and the formation and evolution of superclusters in the local Universe \citep[][]{2024arXiv241208708S}. In this work, we further demonstrate that these simulations also capture details of the individual thermodynamic state of the galaxy clusters in the local Universe. 
 

This ability to replicate galaxy cluster profiles, which are intrinsically linked to their formation histories, suggests that the simulations can reliably narrow the possible evolutionary trajectories of these systems. Previous studies reinforce this interpretation. In particular, \citet{2016MNRAS.460.2015S} demonstrated that Virgo-like halos emerging from constrained initial conditions exhibit more consistent and less stochastic mass accretion histories compared to those found in random simulations. This implies that the formation path of constrained clusters is not arbitrary, but rather shaped by the large-scale structure encoded in the observational constraints. 

Consequently, we establish robust one-to-one connections between the formation history of individual clusters and their thermodynamic states at \(z = 0\), providing crucial insights into their evolutionary pathways. By linking cluster formation processes to their present-day ICM properties, we aim to enhance our understanding of the mechanisms shaping galaxy clusters and refine models of their evolution in a cosmological context.

The paper is structured as follows: Sec.~\ref{sec:Observational data} describes the collected observational X-ray and thermal Sunyaev–Zeldovich (tSZ) data and the methods originally used to extract deprojected thermodynamic profiles from the observations. Sec.~\ref{sec:Simulations} outlines the simulation setup, including the constrained initial conditions and the SLOW simulation suite, as well as the identification of simulated counterparts to observed clusters. Sec.~\ref{sec:Results} presents our results, comparing observed and simulated cluster profiles and assessing their agreement. In Sec~\ref{subsec:cores} we examine the link between formation history and core classification (CC, WCC, NCC). Finally, Sec.~\ref{sec:Conclusions} summarizes our conclusions, discussing the implications for modeling cluster cores and highlighting future improvements in subgrid physics, AGN feedback, and turbulence modeling.

\section{Observational data}
\label{sec:Observational data}
We compiled a robust dataset of observed deprojected profiles by gathering results from the literature \citep[e.g.,][]{2006ApJ...640..691V, 2010A&A...517A..92A, 2013A&A...551A..23E, 2013A&A...550A.131P, 2019A&A...621A..41G}, covering up to 12 Local Universe galaxy clusters. These profiles were originally derived from X-ray observations conducted with the \textit{Suzaku} \citep{2007PASJ...59S...1M}, \textit{Chandra} \citep[][]{2002PASP..114....1W}, and \textit{XMM-Newton} \citep[][]{2001A&A...365L...1J} telescopes.

The XCOP project \citep[][]{2017AN....338..293E} served as a key source,  offering deprojected thermodynamic profiles such as temperature, density, pressure, and entropy for eight galaxy clusters, including A644, A1644, A1795, A2029, A2319, A3158, and A3266 \citep[][]{2019A&A...621A..41G}. These profiles were extracted under the assumption of spherical symmetry and were constructed by combining X-ray observations from XMM-Newton with tSZ measurements from \textit{Planck} 2015 data release \cite[i.e. the full intensity survey, see][Sec. 2 for further details]{2019A&A...621A..39E}. 

In the case of A85 we compared to data presented by \citet{2015MNRAS.448.2971I}. The deprojected thermodynamic profiles of Abell 85 presented by \citet{2015MNRAS.448.2971I} were derived using Suzaku observations under the assumption of spherical symmetry, focusing on the direction of the infall of the southern subcluster (S subcluster). Spectral data were extracted from concentric annuli centered on the cluster and fitted employing the projection model to deproject the overlapping emission along the line of sight. Each shell was modeled as a single-temperature thermal plasma in collisional ionization equilibrium, with temperature and normalization as free parameters. The pressure profile was calculated using the ideal gas law, while the entropy profile was determined from the relation \( S = kT n_e^{-2/3} \). These profiles were compared to theoretical models \citep[e.g.,][]{2007ApJ...668....1N,2010A&A...511A..85P} to validate their accuracy and extend the analysis out to the virial radius (\( r_{200} \)). Projection corrections were also applied to refine the three-dimensional distributions.

The observed deprojected thermodynamic profiles for Abell 119, were presented in Fig.~\ref{A119} of \citet[][]{2010A&A...513A..37H}. They were extracted using radial profiles of X-ray surface brightness from Chandra observations. The temperature, density, and pressure profiles were deprojected assuming spherical symmetry. The X-ray surface brightness was used to infer emissivity and gas density under the assumption of constant density within spherical shells. The deprojected density was combined with the projected temperature to calculate the pressure profile.

The deprojected thermodynamic profiles of the Coma Cluster were obtained from multiple observational datasets, including X-ray and tSZ - effect measurements. The deprojected pressure profile presented in \citet[][Fig. 13]{2013A&A...554A.140P} was derived using SZ data from the \textit{Planck} satellite. 

We compare to azimuthally resolved deprojected profiles presented by \citet[][]{2013ApJ...775....4S}, based on Suzaku X-ray observations, allowing for sector-dependent analyses, that included the E+NE and NW+W directions, showing regional variations in the thermodynamic properties. The deprojected profiles were obtained using the onion-shell deprojection technique, assuming spherical symmetry. Furthermore, we incorporate the median deprojected profiles from \citet{2020MNRAS.497.3204M}, together with the shaded regions representing the spread across all measured sectors.

We include observations of the Perseus cluster, one of the most X-ray luminous galaxy clusters in the local Universe. The deprojected thermodynamic profiles of Perseus were extracted from \citet[][]{2014MNRAS.437.3939U}, which utilized XMM-Newton observations. The study provided electron density, temperature, entropy, and pressure profiles derived under the assumption of spherical symmetry. As usual, the deprojection analysis was performed using an onion-shell technique.

We also include observations of the Virgo cluster, the nearest massive galaxy cluster at a distance of 16 Mpc, making it a key system for detailed structural studies. The deprojected thermodynamic profiles of Virgo were extracted from \citet[][]{2011MNRAS.414.2101U}, which used a mosaic of XMM-Newton observations covering the cluster out to its virial radius. The study found that Virgo has a relatively shallow gas density profile, with a power-law index of \( \beta = 1.21 \pm 0.12 \), indicating a less concentrated gas distribution compared to more massive, relaxed clusters, and presented an entropy profile that follows a gravitational collapse-like power law (\( K \propto r^{1.1} \)) within 450 kpc, but flattens at larger radii, falling below theoretical expectations. As stated by \citet[][]{2011MNRAS.414.2101U}, due to the dynamically unrelaxed nature of Virgo, significant substructure is present and deviations from hydrostatic equilibrium are expected. This unrelaxed state has been independently confirmed by studies of simulated Virgo replicas \citep[][]{2016MNRAS.460.2015S, 2021MNRAS.504.2998S, 2024A&A...682A.157L}. Consequently, the observed profiles are affected by systematic uncertainties arising from departures from spherical symmetry, which are significantly larger than the measurement errors, making direct comparisons with theoretical models challenging. Nevertheless, being aware of these complexities gives us a strong foundation to test the robustness of simulation predictions for unrelaxed clusters.

\section{Simulations}
\label{sec:Simulations}
\subsection{Constrained Initial Conditions}

The initial conditions of the simulations have been described extensively in \citet{2018MNRAS.478.5199S, 2023A&A...677A.169D, 2024A&A...687A.253H}, so we limit the description here to a short overview.

The initial conditions for this study are derived from constrained simulations tailored to replicate the local Universe's large-scale structures. Unlike traditional cosmological simulations, which generate random realizations of the Universe, constrained simulations aim to match observed structures within a specified volume \citep[e.g.,][]{1991ApJ...380L...5H,2010arXiv1005.2687G, 2016MNRAS.455.2078S}. This is achieved by using observational data to set constraints on the initial density and velocity fields used to initialize the simulations. 

For this work, we employed constraints based on peculiar velocities from the CosmicFlows-2 catalog \citep[][]{2013AJ....146...86T}, which contains over 8,000 galaxies with measured distance moduli derived from indicators such as the Tully-Fisher relation, the fundamental plane, and supernovae. Radial peculiar velocities, calculated from these distance measurements and redshifts, are used to reconstruct the three-dimensional velocity field of the local Universe. The main steps within this process are the grouping of the data \citep[e.g.][]{2018MNRAS.476.4362S}, the minimization of biases \citep[e.g.,][]{2015MNRAS.450.2644S}, and the reconstruction of the cosmic displacement field using the Wiener Filter method \citep[][]{1995ApJ...449..446Z} applied to the radial peculiar velocities. This procedure is designed to minimize variance, enhancing the reconstruction's accuracy, while addressing biases and inherent observational noise \citep[][]{1914ApJ....40..187K, 1922MeLuF.100....1M, 1992ApJ...391..494L}.

To relocate the galaxy and group constraints to the positions of their progenitors in the initial density field, the reconstructed cosmic displacement field is used through the Reverse Zel’dovich Approximation \citep[][]{2013MNRAS.430..888D,2014MNRAS.437.3586S}. The Gaussian random field used to create the initial conditions is then set to be consistent with the constraints through the constrained realizations technique \citep[][]{1991ApJ...380L...5H}. This technique, applied at scales where observational data dominates, ensures that the large-scale structure matches observations, while smaller-scale features are populated statistically consistent with the prior $\Lambda$CDM cosmology.

\subsection{Simulation Suite: SLOW}
The SLOW (Simulating the Local Web) simulations\footnote{\url{https://www.usm.lmu.de/~dolag/Simulations/\#SLOW}} have been performed using the realization number 8 of CLONES \citep{2018MNRAS.478.5199S}. It consists of cosmological simulation boxes of \(500 \, h^{-1} \mathrm{Mpc}\) per side, that incorporate the constrained initial conditions described above and use the \textit{Planck} cosmological parameters (\(\Omega_m=0.307\), \(\Omega_\Lambda=0.693\), \(H_0=67.77 \, \mathrm{km \, s^{-1} \, Mpc^{-1}}\), \(\sigma_8=0.829\))  \citep[][]{2014A&A...571A..16P} as their baseline. For the analysis presented in this work, we used the AGN1576$^3$ simulation \citep[see][Table 1]{2024A&A...687A.253H} of the SLOW simulation set.

This simulation has been performed using the OpenGadget3  \cite[OG3 Collaboration in prep.,][and references therein]{2023MNRAS.526..616G} smoothed particle hydrodynamics (SPH) code, an improved version of the GADGET2 code \citep[][]{2005MNRAS.364.1105S}. This code integrates both dark matter and baryonic matter with an SPH formulation \citep{2016MNRAS.455.2110B}. The baryonic physics modules are based on the Magneticum simulation suite \citep[e.g.,][]{2014MNRAS.442.2304H,2015ApJ...812...29T,2016MNRAS.463.1797D, 2025arXiv250401061D} and include prescriptions for star formation, cooling, chemical enrichment, and AGN feedback. Subgrid models for black hole accretion and supernova-driven winds \citep[e.g.,][]{2003MNRAS.339..289S,2010MNRAS.401.1670F,2014MNRAS.442.2304H,2015MNRAS.448.1504S,2004MNRAS.349L..19T,2007MNRAS.382.1050T} ensure a realistic representation of galaxy formation and evolution. For more details, see \citep{2023A&A...677A.169D,2024A&A...687A.253H}.

\subsection{Identification of Galaxy Clusters}

Halos are identified using \textsc{SubFind} \citep{2001MNRAS.328..726S,do09}, which detects halos based on the standard Friends-of-Friends algorithm \citep{1985ApJ...292..371D} and subhalos as self-bound regions around local density peaks within the main halos. The center of halos and subhalos is defined as the position of the particle with the (local) minimum of the gravitational potential. The mass used, $M_{500,200}$, is defined through the spherical overdensity around a halo for the chosen overdensity. 

The identification of galaxy clusters within the SLOW simulation was presented in detail in \citet[][]{2024A&A...687A.253H}. It was carried out using a combination of multiwavelength observational data and simulated properties. Observed clusters were selected from the SZ Cluster Database \footnote{http://szcluster-db.ias.u-psud.fr} and other local group datasets, focusing mainly on systems with \(M_{500} > 2 \times 10^{14} \, \mathrm{M_\odot}\). For each observed cluster, replicas were found in the simulation by searching for halos within a radius of \(45 \, h^{-1} \mathrm{Mpc}\). The quality of the replication was assessed using a null-hypothesis test that included the distance between observed and simulated position, and the mass, temperature, and X-ray luminosity of the replica.

Galaxy clusters from this SLOW simulation have already been used to interpret recent observations from the eROSITA instrument. Among them, the detected bridge between A2667 and A3651 \citep{2024A&A...691A.286D} as well as the Fornax cluster \citep[][]{2025arXiv250302884R}.

\subsection{Analysis of the Intracluster Medium}

The simulation allows the extraction of thermodynamic properties of the ICM, such as temperature, density, entropy, and pressure, at various radii. These profiles are computed directly from the hydrodynamical outputs of the SLOW-AGN1536$^3$ run, the highest resolution simulation in the suite, which includes full galaxy formation physics. We focus our study on clusters for which we have detailed observational data of their deprojected profiles. 

To better represent the deprojected thermodynamic profiles extracted from observations, we divided the simulated clusters into concentric bins spanning radii from 100 kpc to $3.5$ Mpc (where we note that the virial radius lies typically at $\sim1$~Mpc). The lower limit of 100 kpc is set by the resolution of the simulations, ensuring reliable measurements of the ICM properties. Within each bin, we calculated the mass-weighted averages for pressure, temperature, entropy, and electron density for the hot gas component ($T > 10^6$ K), following an approach consistent with methodologies typically adopted in observational analyses \citep[][]{2007ApJ...668....1N, 2009ApJ...705.1129L, 2019A&A...621A..41G}.

We adopt mass weighting consistently across all thermodynamic quantities to ensure a physically self-consistent representation of the ICM’s thermal energy content. 
Although alternative weighting schemes, such as volume weighting for density or spectroscopic-like weighting for temperature \citep[][]{2004MNRAS.354...10M}, may offer closer analogs to specific observational methods, we prioritize consistency and the physical interpretability of mass-weighted quantities. In particular, mass-weighted temperatures describe the actual thermal energy per unit mass, whereas emission-weighted or spectroscopic temperatures tend to be biased toward denser, cooler gas phases that dominate X-ray emission \citep[][]{2018A&A...618A..39R}.

The use of mass-weighted temperatures is further justified within $r_{500}$, where the gas distribution is more homogeneous and the impact of multiphase structures and clumping is reduced, especially in relaxed clusters. In these regions, the mass-weighted and spectroscopic temperatures generally agree within 10\% to 15\% \citep[e.g.,][]{2004MNRAS.354...10M, 2005ApJ...618L...1R}, which includes intermediate radial ranges that constitute the main focus of our analysis. 

Nevertheless, we recognize that mass-weighted temperatures may overestimate observationally derived temperatures in more complex environments. Larger discrepancies are expected in non-virialized clusters, where substructures and multiphase gas become more prominent, as well as in the innermost regions ($r \lesssim 0.1r_{500}$) where cool cores can bias X-ray measurements towards lower values \citep[][]{1998ApJ...504...27M, 2005ApJ...628..655V}. Similarly, in the outskirts (beyond $r_{500}$), clumping and non-equilibrium effects introduce additional uncertainties. 

An important distinction between our simulation-based profiles and observational studies is that we do not mask substructures in our analysis. X-ray observational studies typically excise or mask substructures, such as infalling clumps and merging sub-halos, to minimize contamination in the derived profiles \citep[][]{2005ApJ...628..655V, 2007A&A...461...71P, 2015Natur.528..105E}. In contrast, by including all gas phases and substructures, our profiles capture the full complexity of the ICM but also introduce greater scatter, particularly in the outskirts and dynamically active systems. To account for this, we present both mean and median profiles. The median provides a robust characterization of the bulk ICM, being less sensitive to high-density outliers. This is a desirable feature given that the distributions of thermodynamic quantities such as pressure and density are often approximately log-normal. In such skewed distributions, the median is generally more representative of the typical value than the mean, which can be significantly biased by dense clumps and shocks. 
The mean, on the other hand, incorporates the contribution from dense clumps and shocks, providing valuable insight into the aforementioned total gas distribution and the level of gas inhomogeneity. The difference between the mean and median serves as a useful diagnostic of the cluster's dynamical state and clumping degree \citep[][]{2013MNRAS.429..799V, 2013MNRAS.428.3274Z}.

To assess the uniqueness and accuracy of each constrained replica, we constructed a control sample of randomly selected clusters from the simulation. For each observed system, we selected up to 50 halos whose masses fall within the range defined by the observationally inferred mass and the mass of the corresponding replica, as reported by \citet{2024A&A...687A.253H}. To define this range, we rounded both bounds to the highest and lowest $10^{14}\,M_\odot$; for example, a mass range from $3.4$ to $2.2 \times 10^{14}\,M_\odot$ would be rounded to $4$ and $2 \times 10^{14}\,M_\odot$, respectively. In case it was not possible to select 50 clusters due to the high mass of the system or when the observational and simulated masses were very similar, we included all available halos from the simulation volume that satisfied the mass criterion.

For each selected halo, we computed the thermodynamic profiles (pressure, temperature, entropy, and electron density) in the same radial bins used for the replicas. We then calculated the full range of values spanned by the sample at each radial bin, as well as the 1$\sigma$ region around the median. These bands provide a quantitative baseline against which to evaluate the performance of the constrained replicas and help identify features that are not reproduced by randomly selected halos of similar mass.

Despite its simplicity, this approach enables a direct and meaningful comparison between simulated and observed deprojected profiles. By focusing on the simulated counterparts of observed clusters, we can directly compare absolute values and radial distances, without the need to scale by $R_{500}$. This strategy makes our comparison less sensitive to errors induced by the hydrostatic mass bias. Indeed, when observational profiles are scaled by $R_{500}$ or $M_{500}$, the derived scaling radius typically assumes hydrostatic equilibrium; if non-thermal pressure support or bulk motions are significant, this can lead to an underestimation of $R_{500}$ and consequently shift the scaled profiles. By comparing profiles in physical units (e.g., absolute distances in kpc), we reduce sensitivity to these biases, although some implicit effects may remain through observational sample selection and analysis choices.\footnote{Note that the masses from the simulation, when compared to the observationally inferred ones, are compatible with a hydrostatic bias of $(1-b) \approx 0.87$; ~\citep[see][]{2024A&A...687A.253H}.}

An important consideration in our comparison between simulated and observational profiles is the cosmological model adopted in each case. Some observational datasets employed in this work are based on different cosmological parameter values, which can introduce systematic differences in key quantities such as radius, electron density, pressure, and entropy. While our simulations adopt the \textit{Planck} cosmology with \(h = 0.67\), \(\Omega_M = 0.315\), and \(\Omega_\Lambda = 0.685\), several observational studies assume a flat \(\Lambda\)CDM cosmology with \(H_0 = 70\ \text{km}\ \text{s}^{-1}\ \text{Mpc}^{-1}\) (\(h = 0.7\)), \(\Omega_M = 0.3\), and \(\Omega_\Lambda = 0.7\). To ensure consistency, we rescale observational quantities that are sensitive to the underlying cosmology. Physical radii scale inversely with the Hubble parameter, \(R \propto h^{-1}\), resulting in an increase of approximately 4.5\% when converting observational radii from \(h=0.7\) to \(h=0.67\). Electron densities, derived from X-ray surface brightness profiles and dependent on the angular diameter distance, scale as \(n_e \propto h^{1/2}\), decreasing by about 2.3\% in this conversion. Pressure, defined as \(P_e = n_e k_B T\), follows the same scaling as density, \(P_e \propto h^{1/2}\), since X-ray temperatures are cosmology-independent. Entropy, given by \(K \propto T n_e^{-2/3}\), rescales as \(K \propto h^{-1/3}\).

We note that the XCOP sample adopts the same \textit{Planck} cosmology as our simulations. Consequently, no rescaling is necessary for the XCOP profiles, which can be directly compared to our simulated clusters. For all other observational data sets, we apply the corresponding cosmology rescalings to enable a consistent and meaningful comparison of the thermodynamic profiles and radial distances.

The constrained nature of the simulations enables a direct connection between the thermodynamic properties of the ICM and the formation histories and environments of individual galaxy clusters. This framework allows us to explore how cluster dynamics and large-scale environmental factors influence the diversity of ICM profiles observed at $z = 0$. By comparing simulated analogs to well-studied observed clusters, we gain valuable insights into the role of mergers, accretion, and surrounding structure in shaping the thermodynamic properties of the ICM and driving cluster-to-cluster variations.

\section{Local Galaxy Cluster Profiles and their Simulated Replicas}
\label{sec:Results}

In the following, we present a detailed analysis of the thermodynamic state of each of the clusters analyzed and the quality of their profile reproduction.

\subsection{The Perseus Cluster}

\renewcommand{\labelitemi}{\textbullet}

\begin{figure*}
\centering
    \includegraphics[width=0.46\linewidth]{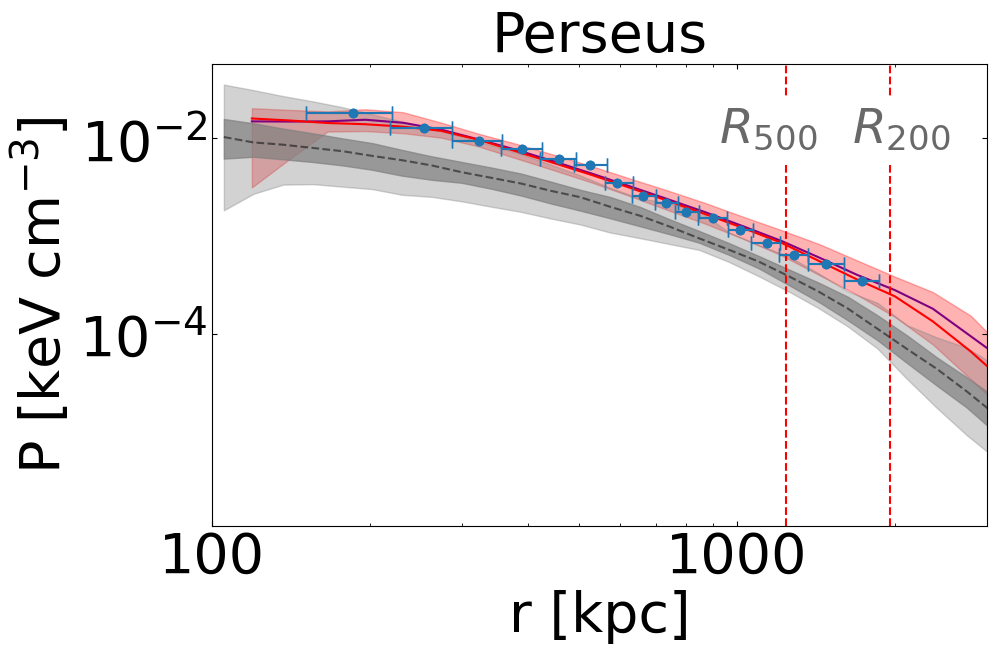}
    \includegraphics[width=0.44\linewidth]{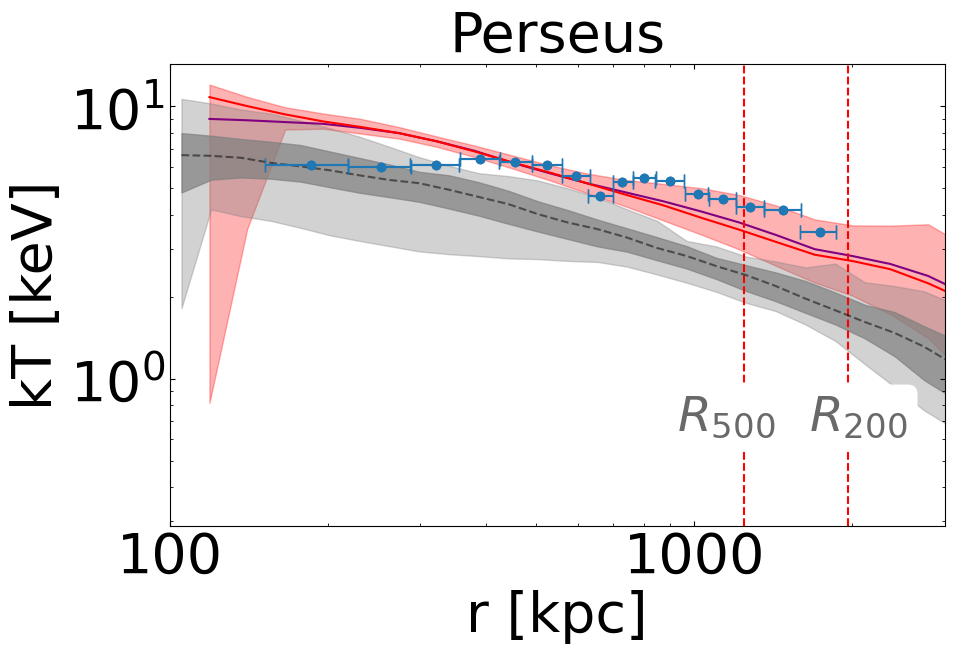}
    \includegraphics[width=0.44\linewidth]{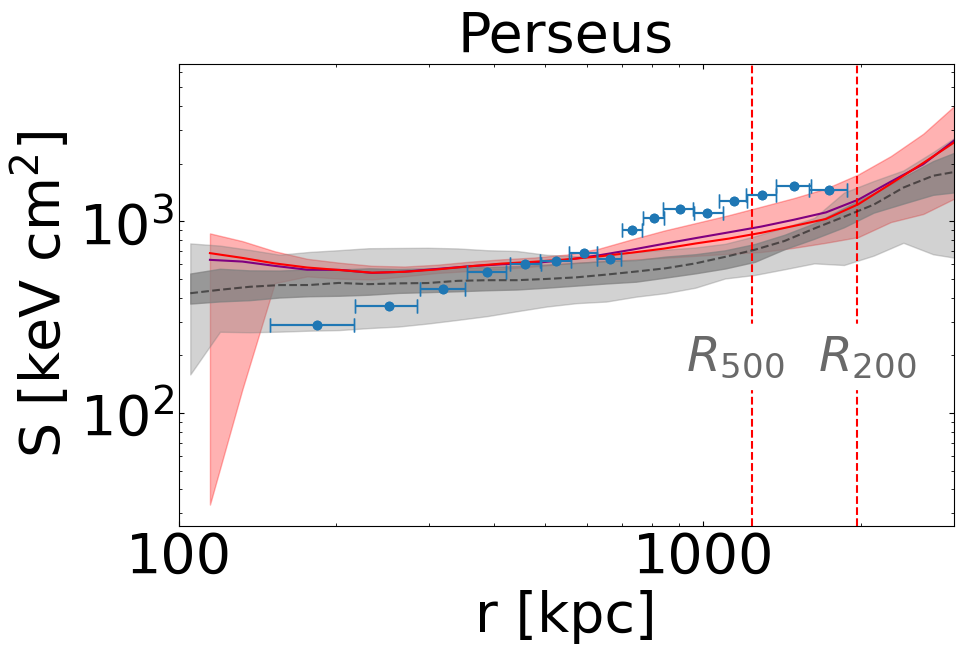}
    \includegraphics[width=0.46\linewidth]{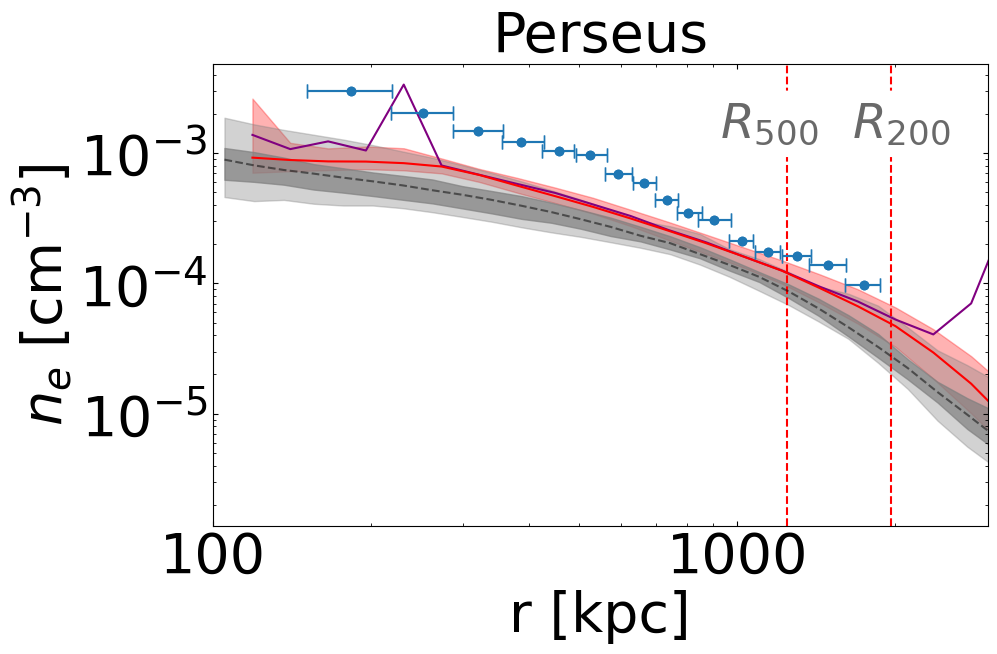}
    \caption{Deprojected thermodynamic profiles of the Perseus cluster compared to observational data from \citep{2014MNRAS.437.3939U}. The panels, from left to right, show the pressure, temperature, entropy, and electron density profiles. The red line represents the median profile of the simulated Perseus cluster, while the red shaded region indicates the \(1\sigma\) scatter of particles within each spherical-shell bin. The purple line represents the mean profile of the simulation. The blue points correspond to deprojected observational data from \citet[][]{2014MNRAS.437.3939U}. The gray line represents the median profile of the random cluster sample; the dark gray shaded region denotes the corresponding $1\sigma$ spread, and the light gray region shows the full range covered by the sample at each radial bin. The vertical dashed lines mark \( r_{500} \) and \( r_{200} \) of the simulations. 
    }

    \label{Perseus}
\end{figure*}

\begin{figure*}
\centering
    \includegraphics[width=0.46\linewidth]{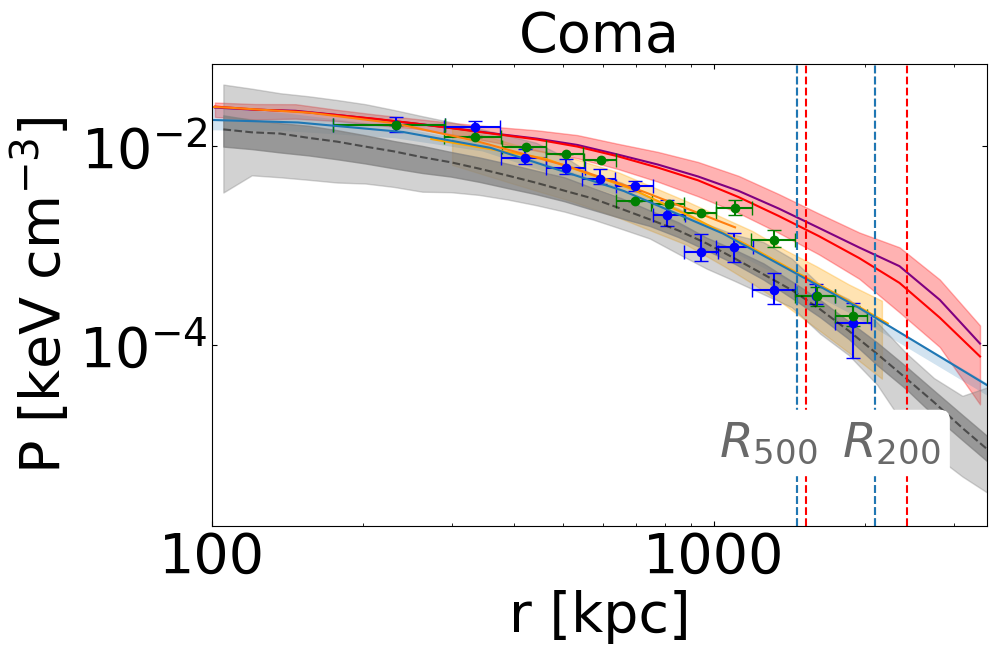}
    \includegraphics[width=0.44\linewidth]{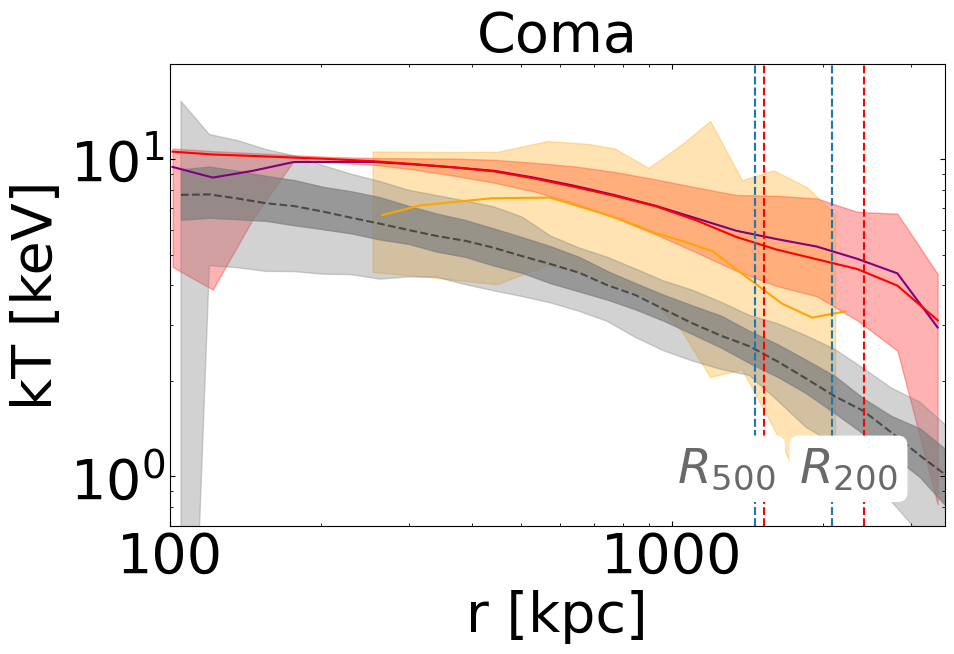}
    \includegraphics[width=0.44\linewidth]{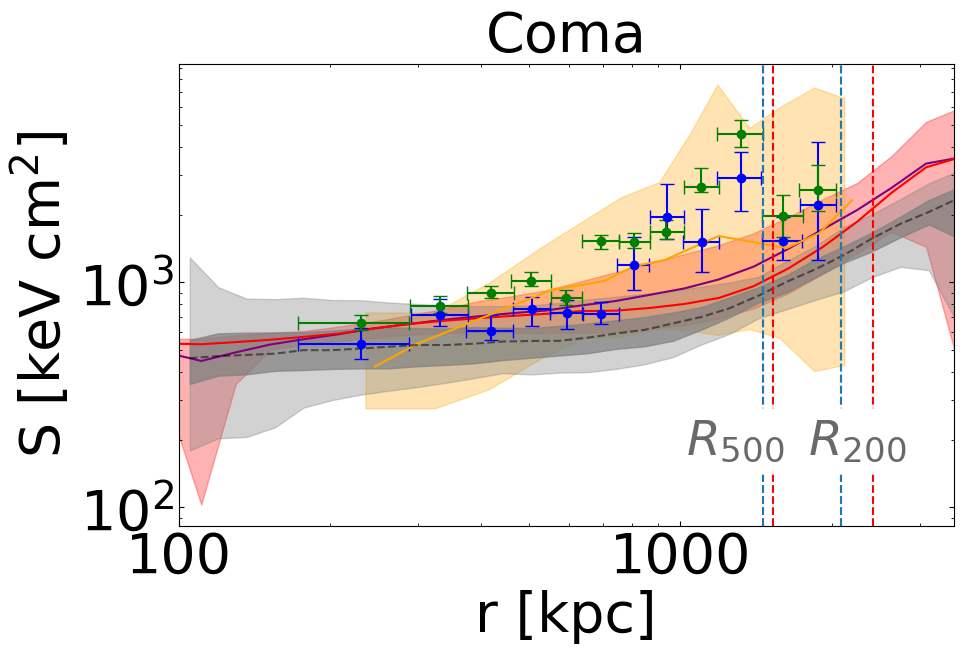}
    \includegraphics[width=0.46\linewidth]{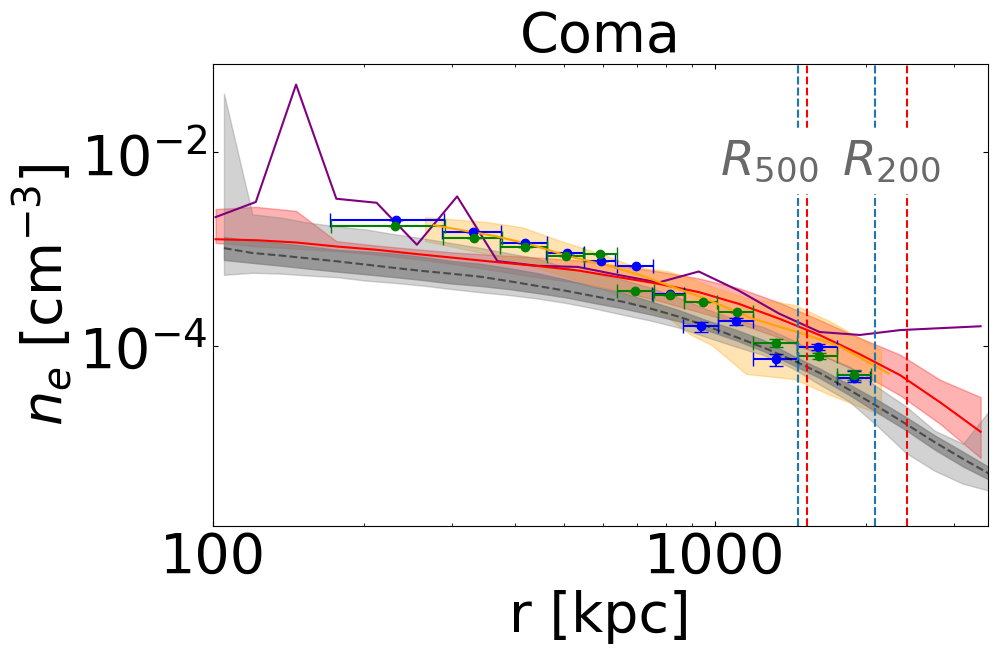}
    \caption{Deprojected thermodynamic profiles of the Coma cluster, comparing simulation results to observational data. The panels, from left to right, show the pressure, temperature, entropy, and electron density profiles. The red line represents the median profile of the simulated Coma cluster, while the red shaded region indicates the \(1\sigma\) scatter of particles within each spherical-shell bin. The purple line represents the mean profile of the simulation. Observational data from A. \citet[][]{2013ApJ...775....4S} are shown as blue points (E+NE direction) and green points (NW+W direction). The yellow line represents the median deprojected pressure profile from \citet[][]{2020MNRAS.497.3204M}, with the shaded yellow region indicating the range covered by profiles in all observed sectors. The gray line represents the median profile of the random cluster sample; the dark gray shaded region denotes the corresponding $1\sigma$ spread, and the light gray region shows the full range covered by the sample at each radial bin. In the leftmost panel, the light blue line represents the pressure profile derived from \textit{Planck} data as presented by \citet[][]{2013A&A...554A.140P}. The vertical dashed lines indicate \( r_{500} \) and \( r_{200} \) from observations (blue) and simulations (red). 
    }
    \label{Coma}
\end{figure*}

Perseus is a well-studied SCC cluster and one of the most X-ray luminous clusters in the local Universe \citep[e.g.,][]{2000MNRAS.318L..65F,2003ApJ...590..225C}. Generally, observations reveal a well-developed cool core with a steep temperature gradient and an entropy profile that flattens at small radii, consistent with a cooling flow region \citep[][]{2014MNRAS.437.3939U}. The pressure and density profiles indicate a relatively relaxed structure, with no major recent disturbances within \( r_{500} \) \citep[][]{2015MNRAS.450.4184Z}.  

Figure \ref{Perseus} presents deprojected thermodynamic profiles of Perseus from \citet[][]{2014MNRAS.437.3939U} together with the profiles of our simulated replica.

\begin{itemize}
\item {\bf The pressure profile}  is shown in the upper left panel of Fig.~\ref{Perseus}. The blue points correspond to the observational data from \citet{2014MNRAS.437.3939U}, while the red line represents the median pressure profile of the simulated Perseus analog. The red shaded region indicates the 1$\sigma$ scatter of gas particles within each spherical shell. The simulated profile closely follows the observed data, with the narrow 1$\sigma$ region suggesting a smooth and well-converged pressure distribution in the simulated intracluster medium (ICM).

For comparison, the gray line shows the median pressure profile of 50 randomly selected clusters from the simulation volume with masses similar to Perseus. The light-gray shaded area encompasses the full range of values spanned by these profiles, while the dark-gray region represents their 1$\sigma$ scatter. Notably, none of the randomly selected clusters reproduces the observed profile as accurately as the Perseus analog. From radii beyond 200 kpc, the Perseus replica's median profile lies outside the 1$\sigma$ region of the random sample, highlighting it as an outlier. Among the simulated clusters, only the constrained Perseus replica matches the observed data this closely.

\item {\bf The temperature profile}  is shown in the upper right panel of Fig.~\ref{Perseus}. The observational data exhibit a rise from the cluster core, reaching a peak at intermediate radii (r $\sim400$ kpc). The temperature profile of the simulated Perseus analog reproduces the observed temperature beyond r $\sim400$ kpc remarkably well, especially when compared to the random cluster sample. At all radii, the simulated profile lies outside the gray-shaded region representing the random sample, underscoring the distinctiveness of the constrained replica. Beyond $\sim400$ kpc, the agreement with observations is particularly strong.

At smaller radii ($r \lesssim 200$ kpc), however, the simulation slightly overestimates the temperature compared to the observed data. This discrepancy may point to differences in the efficiency of cooling processes or the implementation of AGN feedback, both of which are known to play a significant role in shaping the thermodynamic structure of cluster cores \citep[e.g.,][]{2006MNRAS.366..417F, 2007ARA&A..45..117M}.

\item {\bf The entropy profile} is shown in the lower left panel of Fig.~\ref{Perseus}. Astrophysical entropy provides a diagnostic of departures from self-similarity in the intracluster medium (ICM), primarily driven by non-gravitational processes such as radiative cooling, star formation, and feedback from supernovae and active galactic nuclei (AGN). As a result, the entropy profile is highly sensitive to the subgrid physics implemented in the simulation.

The observational data for Perseus display the characteristic flattening of the entropy profile at small radii, a signature of cooling flow clusters \citep[][]{2009ApJS..182...12C, 2013Natur.502..656W}. The simulated Perseus replica reproduces this trend qualitatively, though it modestly overestimates the central entropy, likely reflecting limitations or uncertainties in the modeling of feedback processes.

While both the replica and the random sample struggle to fully reproduce the observed shape of the entropy profile in the core, the Perseus analog more closely tracks the observed data across a large portion of the radial range, particularly beyond 250 kpc. This underscores the value of constrained simulations and one-to-one comparisons with observed clusters, as they offer a powerful means to evaluate and refine feedback models to better capture the thermodynamic structure of systems like Perseus.

\item {\bf The electron density profile} is displayed in the lower right panel of Fig.~\ref{Perseus}. The profile of the simulated Perseus analog lies in the upper envelope of the random cluster sample, making it the closest match to the observational data among all simulated clusters with similar mass. However, it still falls short of reproducing the exact electron densities observed for Perseus, particularly at smaller radii.

This discrepancy may, in part, be attributed to resolution limitations, which are known to significantly affect the accuracy of density measurements in the cluster core. Higher resolution simulations could potentially lead to improved agreement with the observed profile (see Appendix \ref{appendix1} for a discussion on this topic). Nevertheless, among the available simulated clusters, the selected Perseus replica provides the best overall match to the observed electron density distribution.

\end{itemize}

\begin{figure*}
  \centering

    \includegraphics[width=0.46\linewidth]{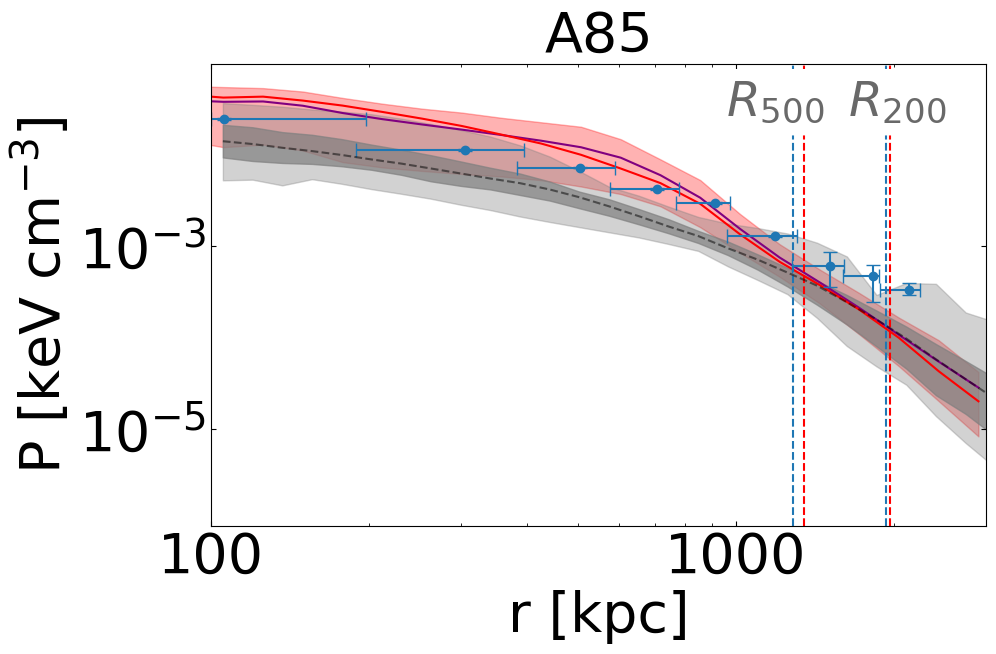}
    \includegraphics[width=0.44\linewidth]{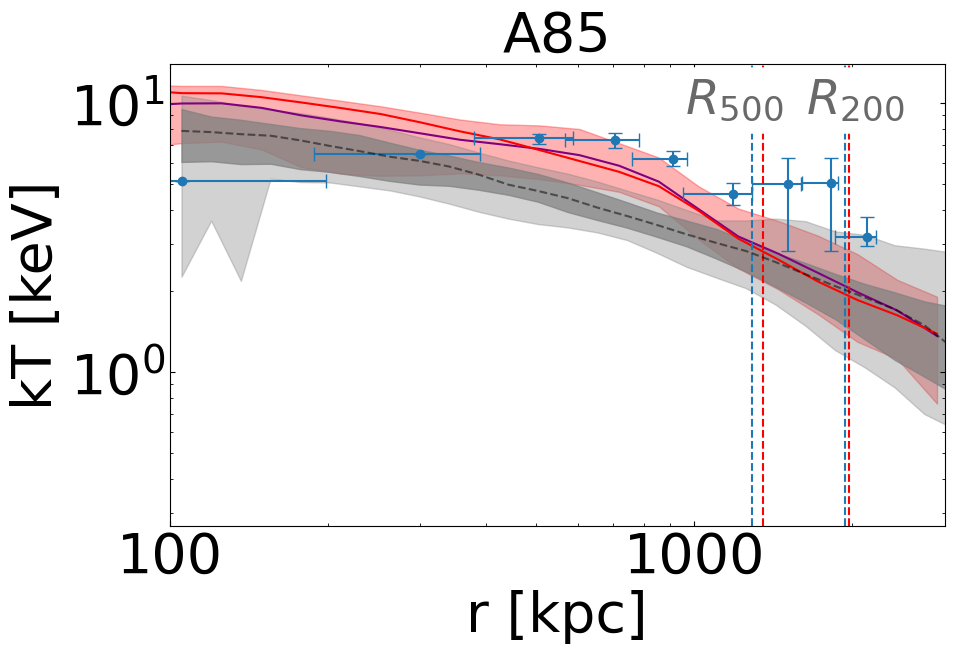}
    \includegraphics[width=0.44\linewidth]{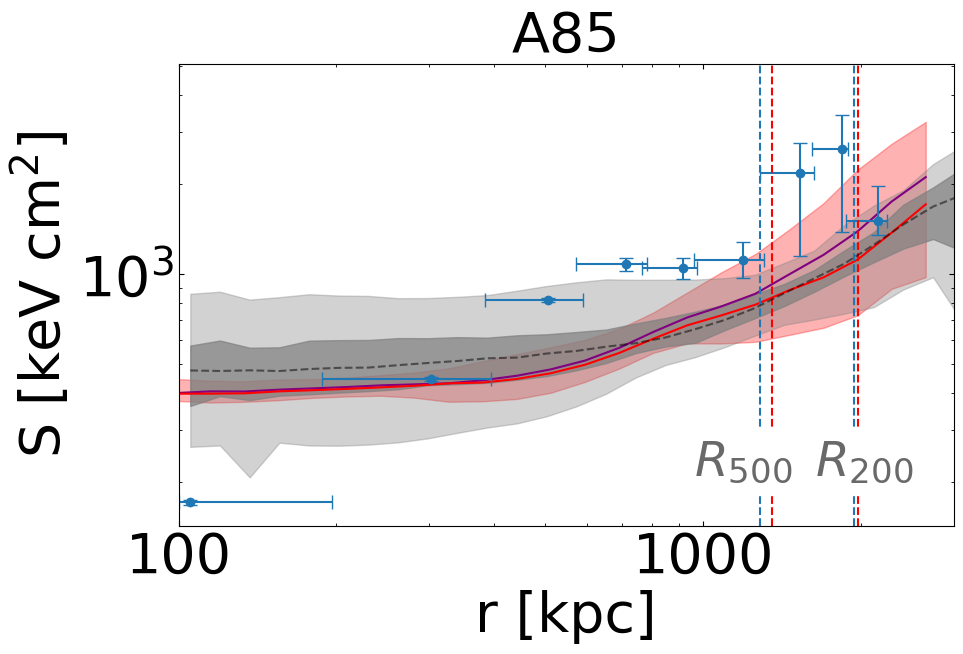}
    \includegraphics[width=0.46\linewidth]{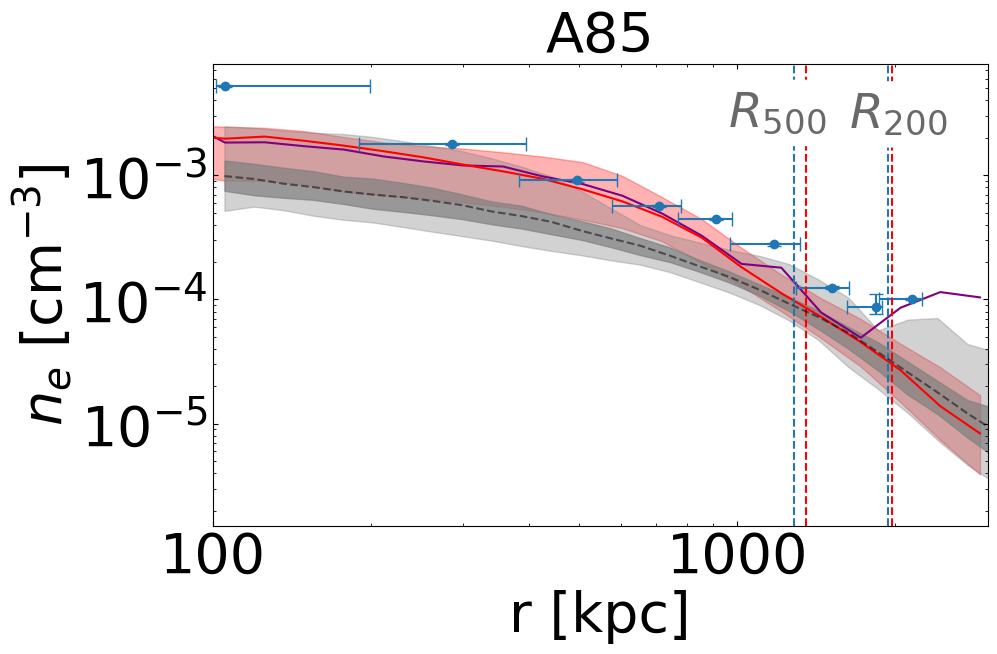}
    \caption{Deprojected thermodynamic profiles of the galaxy cluster Abell 85. The panels show (from left to right) the pressure, temperature, entropy, and electron density profiles as a function of radius. The red solid line represents the median of the simulated cluster, with the red shaded region indicating the $ \sigma $ scatter within each radial bin. The purple line represents the mean profile of the simulation.  The gray line represents the median profile of the random cluster sample; the dark gray shaded region denotes the corresponding $1\sigma$ spread, and the light gray region shows the full range covered by the sample at each radial bin. The blue data points correspond to deprojected observational data extracted using the Suzaku telescope and presented by Ichinohe et al. (2015). The vertical dashed lines indicate $ r_{500} $ and $ r_{200} $ from observations. 
    }
    \label{A85}
\end{figure*} 

\begin{figure*}
  \centering
    \includegraphics[width=0.45\linewidth]{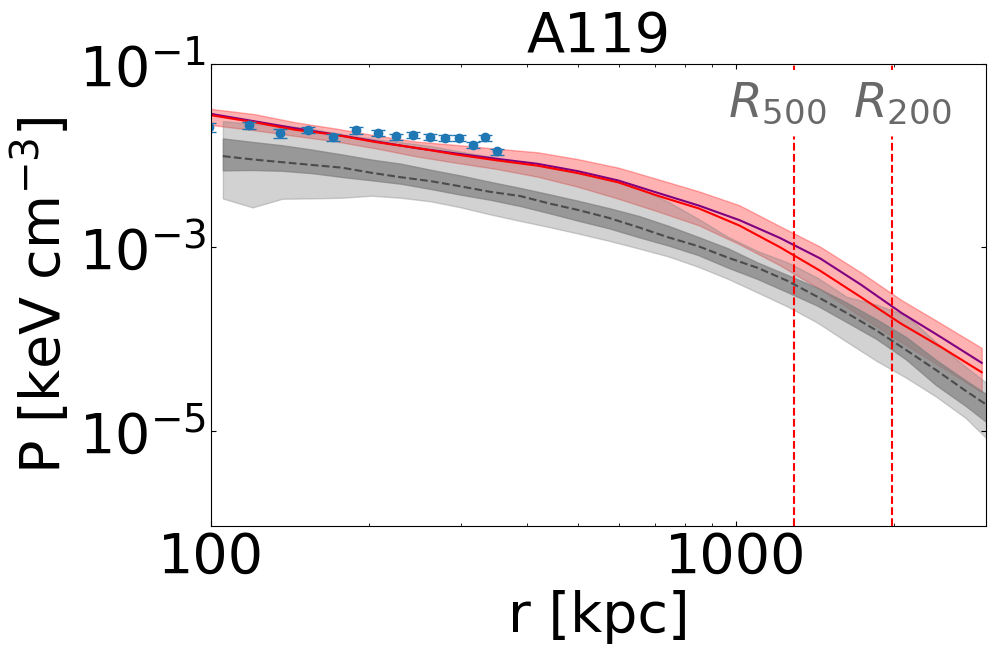}
    \includegraphics[width=0.45\linewidth]{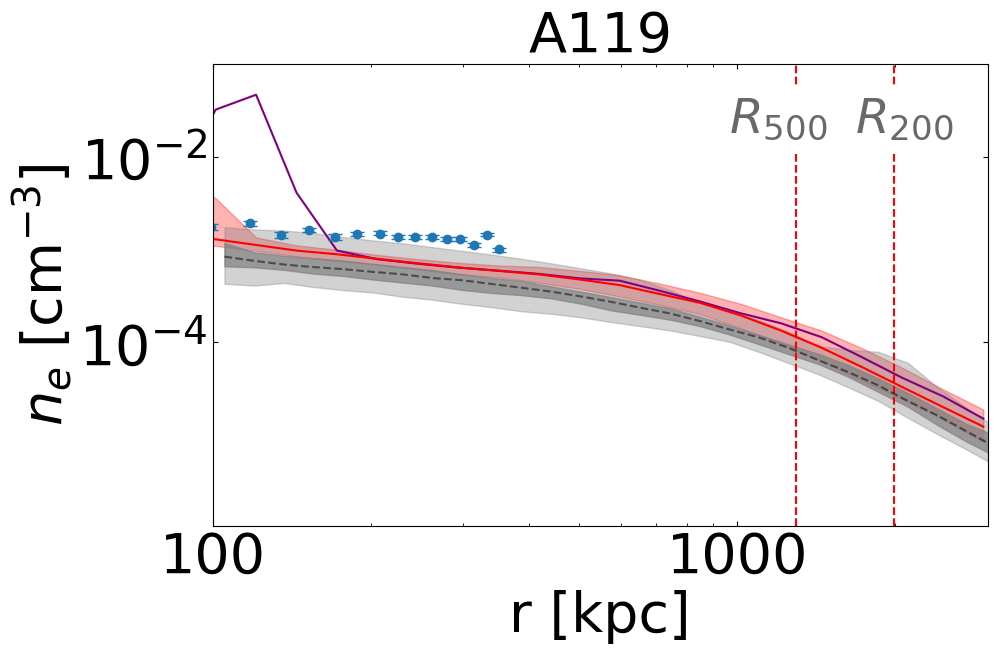}
    \caption{Deprojected thermodynamic profiles of the galaxy cluster Abell 119. The panels show (from left to right) the electron density and pressure profiles as a function of radius. The red solid line represents the median of the simulated cluster, with the red shaded region indicating the $ \sigma $ scatter within each radial bin. The purple line represents the mean profile of the simulation.  The gray line represents the median profile of the random cluster sample; the dark gray shaded region denotes the corresponding $1\sigma$ spread, and the light gray region shows the full range covered by the sample at each radial bin. The blue data points correspond to deprojected Chandra observations from the ACCEPT database. The vertical dashed lines mark characteristic radii, including $ r_{500} $ and $ r_{200} $ of the simulation. The observation profiles assume spherical symmetry and are obtained using an onion-shell deprojection technique. }
    \label{A119}
\end{figure*}

\subsection{The Coma Cluster}

The Coma Cluster is a massive, dynamically evolved, and nearly isothermal galaxy cluster, often used as a reference system for studying the thermodynamic properties of the ICM  \citep[][]{1992A&A...259L..31B,2001A&A...365L..67A,2013ApJ...775....4S}. Unlike SCC clusters, Coma exhibits a flat central entropy profile and lacks a strongly peaked density distribution, characteristics typical of NCC systems \citep[][]{2009ApJS..182...12C,2014MNRAS.438..195P}. Observations indicate that the ICM is well mixed, with no evidence of a dominant central cooling flow \citep[][]{2013PASJ...65...10M}.

Figure \ref{Coma} presents the deprojected thermodynamic profiles of Coma from multiple observational datasets compared to simulated profiles. Observational measurements include deprojected profiles from \citet[][]{2013ApJ...775....4S}, where green points represent the NW+W direction and blue points represent the E+NE direction. Additionally, the yellow line represents the median deprojected pressure from \citep[][]{2020MNRAS.497.3204M}, with the shaded region indicating the range covered by deprojected profiles across all sectors. In the leftmost panel, the light blue line represents \textit{Planck} data. 

\begin{itemize}
\item {\bf The pressure profile}  is shown in the upper left panel of Fig.~\ref{Coma}. The observational median profile reported by \citet[][]{2020MNRAS.497.3204M} closely follows the median of the random cluster sample. However, several individual data points extracted from the NW+W sector (green points) align well with the profile of the simulated Coma analog, particularly at intermediate and large radii.

\item {\bf The temperature profile} is presented in the upper right panel of Fig.~\ref{Coma}. The observed median profile from \citet[][]{2020MNRAS.497.3204M} remains relatively flat, consistent with the Coma Cluster’s well-known isothermal nature \citep[][]{2001A&A...365L..67A}. In contrast, the random sample of simulated clusters exhibits a clear declining trend with radius, which does not reflect the observed behavior. The Coma analog, however, successfully reproduces the flat temperature distribution, maintaining a nearly constant temperature across a broad radial range and demonstrating a closer match to the observations.

\item {\bf The entropy profile} is shown in the lower left  panel of Fig.~\ref{Coma}. Despite the inherent uncertainties associated with astrophysical processes affecting entropy in the ICM, the simulated Coma analog reproduces the observed median profile reported by \citet[][]{2020MNRAS.497.3204M} within its 1$\sigma$ region from approximately 400 kpc out to the cluster outskirts. In contrast, the random cluster sample systematically underestimates the entropy across most radii, highlighting the improved agreement achieved through the constrained simulation.

\item {\bf The electron density profile}  is shown in the lower right panel of Fig.~\ref{Coma}. As a non-cool-core (NCC) system, Coma features a more diffuse central gas distribution, which means that the simulation is less affected by the resolution-driven underestimation of central densities often seen in simulated cool-core clusters. The yellow line from \citet[][]{2020MNRAS.497.3204M} represents the sector-averaged observational profile, with the simulated Coma analog remaining well within the observed spread across all radii.
\end{itemize}

However, the difference between the random sample and the replica is relatively small, limiting the strength of any conclusions that can be drawn based on the density profile alone. Nonetheless, in the case of the Coma analog, we find a significant deviation between the mean and the median values in the cluster core, indicating the presence of substantial substructure, consistent with the known dynamical complexity of the observed Coma Cluster.

\subsection{Abell 0085}
Abell 0085 (also known as A85) is a well-known SCC cluster characterized by its steep central temperature gradient, well-developed core, and relatively relaxed dynamical state  \citep[][]{2002ApJ...567..716R}. Suzaku observations have provided detailed measurements of the cluster’s electron density, temperature, pressure, and entropy profiles, which have revealed the thermodynamic properties of the ICM of A85 across a wide range of radii \citep[][]{2010ApJ...714..423K, 2015MNRAS.448.2971I}.

\begin{itemize}
\item {\bf The pressure profile}  of A85 is shown in the upper left panel of Fig.~\ref{A85}. The observational data points lie outside the 1$\sigma$ region of the random cluster sample at intermediate radii (r $\sim 500-800$ kpc), indicating that typical clusters of similar mass do not reproduce the observed profile. In contrast, the median profile of the A85 simulated replica closely follows the observed data over a broad radial range, from approximately 100 kpc to 1000 kpc, demonstrating the improved agreement achieved through the constrained simulation.

\item {\bf The temperature profile} is shown in the upper right panel of Fig.~\ref{A85} and exhibits a similar trend. Most of the observational data points lie outside the 1$\sigma$ region of the random sample, indicating that the typical simulated clusters fail to capture the observed temperature distribution. In contrast, both the median and mean profiles of the A85 simulated replica closely follow the observed data, showing a temperature peak at $r \approx 600$ kpc followed by a decline at larger radii.

\item {\bf The entropy profile} shown in the lower left panel of Fig.~\ref{A85}, reveals similar challenges to those noted for the other clusters. Both the A85 simulated replica and the random cluster sample struggle to reproduce the overall shape of the observed entropy profile. While the replica may offer a closer match in certain radial ranges, especially at $r > 1000$ kpc, the general trend remains difficult to capture accurately, likely reflecting limitations in the modeling of subgrid physics and feedback processes that strongly influence entropy in the core and outskirts.

\item {\bf The electron density profile}, shown in the upper right panel of Fig.~\ref{A85}, further highlights the accuracy of the simulated A85 replica. Across all radial bins, most of the observed data points lie outside the 1$\sigma$ region of the random cluster sample, indicating that typical clusters in the simulation do not reproduce the observed density structure. In contrast, both the mean and median profiles of the A85 replica closely follow the observed values between $\sim300$ kpc and 1000 kpc. 
\end{itemize}

\subsection{Abell 0119}
Abell 0119 (also known as A119) is considered a NCC cluster that exhibits significant elongation toward the northeast, suggesting it may be undergoing a merger \citep[][]{2004IJMPD..13.1549G,2010A&A...513A..37H,2015A&A...575A.127P}. The cluster is highly diffuse, lacking a prominent cool core, and features a large core region with slowly varying surface brightness, which makes deprojection inaccurate, as spherical symmetry is not a good approximation for the geometry of this cluster. The electron density and pressure profiles of A119, shown in Fig.~\ref{A119}, were extracted from \citet[][]{2009ApJS..182...12C}  using {\it Chandra} observations and extend up to 500 kpc from the cluster's center.

\begin{itemize}
\item {\bf The pressure profile}, shown in the left panel of Fig.~\ref{A119}, demonstrates that the observed profile of A119 lies near the upper boundary of the 1$\sigma$ region of the random cluster sample, remaining outside it across most radii. In contrast, the pressure profile of the simulated A119 replica successfully reproduces the observed values, closely tracking the data and providing a significantly better match than the typical clusters in the simulation volume.

\item {\bf The electron density profile} is shown in the left panel of Fig.~\ref{A119}. The values reported by \citet[][]{2009ApJS..182...12C} are systematically higher than those of both the simulated A119 replica and the random cluster sample. This discrepancy may be partially attributed to the diffuse and highly disturbed nature of A119, which increases the likelihood that the gravitational center used in the simulation does not coincide with the X-ray peak center adopted in the observational analysis. In addition, the pronounced asymmetry of the cluster structure renders the assumption of spherical symmetry problematic for both deprojection and 3D binning, introducing further uncertainty and potential biases in the comparison \citep[][]{2008ApJ...685..118V}.
\end{itemize}
Despite these challenges related to post-processing and comparison methodologies, the thermodynamic profiles of A11, when taken as a whole, are nonetheless well reproduced by the simulated replica.

\subsection{Abell 0644}

\begin{figure}[ht!]
\centering
    \includegraphics[width=0.95\linewidth]{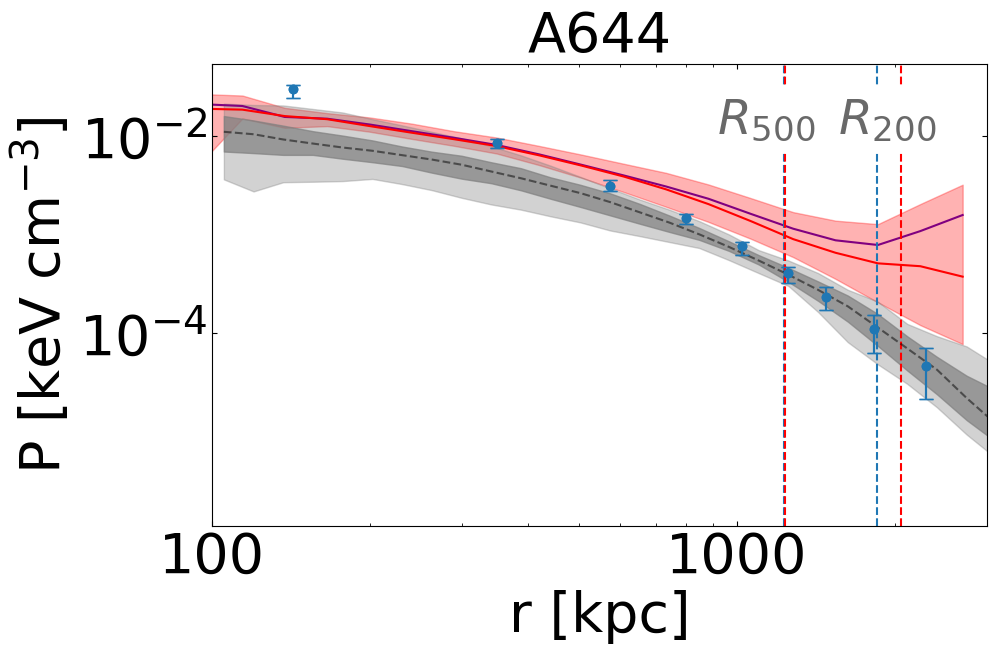}
    \includegraphics[width=0.95\linewidth]{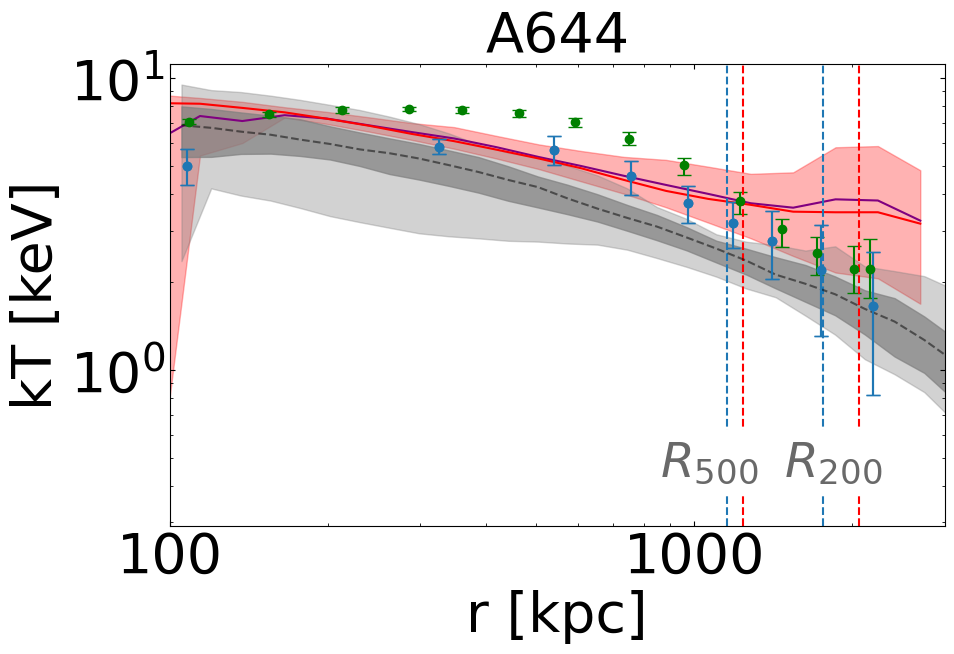}
    \includegraphics[width=0.95\linewidth]{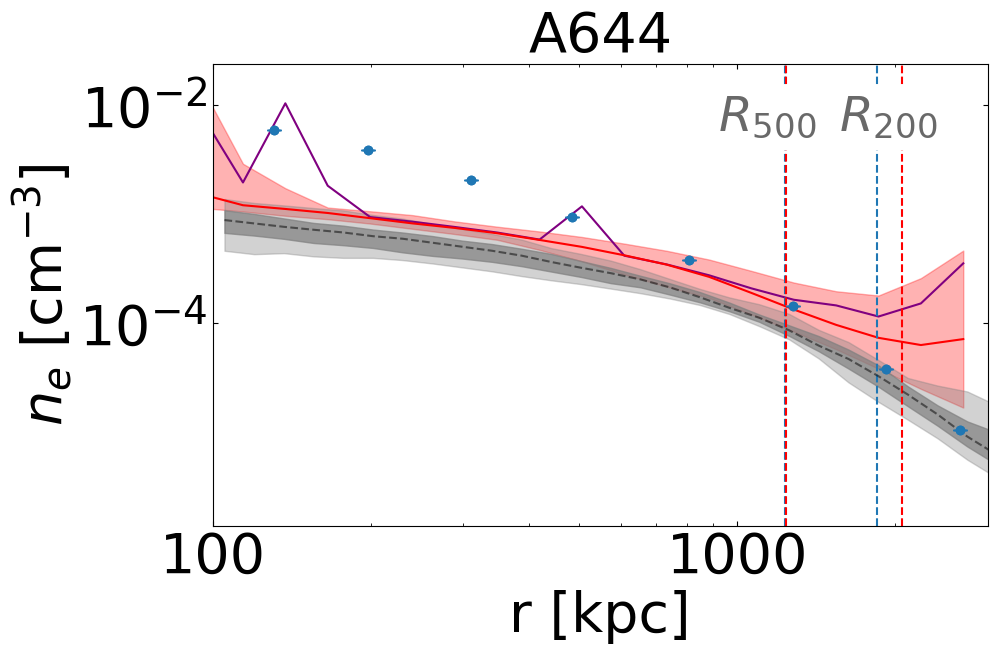}
    \caption{Deprojected thermodynamic profiles of the galaxy cluster Abell 644. The panels show (from left to right) the pressure, temperature, and electron density profiles as a function of radius. The red solid line represents the median of the simulated cluster, with the red shaded region indicating the 1$\sigma$ scatter within each radial bin. The purple line represents the mean profile of the simulation.  The gray line represents the median profile of the random cluster sample; the dark gray shaded region denotes the corresponding $1\sigma$ spread, and the light gray region shows the full range covered by the sample at each radial bin. The blue data points correspond to deprojected XMM-Newton observations from \citet[][]{2019A&A...621A..39E}, while the green points in the temperature panel represent \textit{Planck} data as presented in the same study. The vertical dashed lines mark characteristic radii, including $r_{500}$ and $r_{200}$. 
    }
    \label{A644}
\end{figure}

Abell 0644 (also A644) is a dynamically disturbed galaxy cluster, often categorized as an NCC system \citep[][]{1996ApJ...458...27B,2010ApJ...721L..82C,2011A&A...532A.123R}. Previous studies have revealed significant deviations from symmetry in its ICM, likely driven by recent merger activity. Despite this, the cluster’s large-scale properties remain consistent with relaxed systems \citep[][]{2010A&A...513A..37H}, presenting a unique combination of features. Observations indicate a relatively flat temperature profile in the central regions, reflecting the absence of a strong cool core  \citep[][]{2009ApJS..182...12C,2019MNRAS.483..540L}. This combination of dynamical disturbance and large-scale quasi-relaxation makes A644 an interesting system for studying the interplay between mergers and the evolution of thermodynamic properties in the ICM.

\begin{itemize}
\item {\bf The pressure profile} comparison is shown in the upper panel of Fig.~\ref{A644}. The overall trend observed in the data is not well reproduced by either the A644 simulated replica or the random cluster sample. Within the inner 600 kpc, the observational data points lie outside the 1$\sigma$ region of the random sample but are somewhat better aligned with the profile of the replica. At larger radii, however, the observations converge toward the mean of the random sample, suggesting that while the replica captures some features of the inner pressure distribution, it does not offer a consistently improved match across all scales.

\item {\bf The temperature profile} s shown in the central panel of Fig.~\ref{A644}. The observational data include both temperatures derived from X-ray observations alone (blue points) and from a combination of X-ray and *Planck* data (green points). The X-ray-only profile shows good agreement with the simulated A644 replica at intermediate radii ($r \sim 400\text{–}1000$ kpc), lying outside the 1$\sigma$ region of the random sample. At larger radii, the observed temperatures gradually decline toward the median of the random sample.

The X-ray + {\it Planck} temperature profile yields systematically higher values at intermediate radii. While the simulated replica underestimates these values, it still provides a significantly better match than any of the profiles from the random sample. This suggests that the constrained replica captures the thermal structure of A644 more effectively, particularly in the range where the observational constraints are strongest.

\item {\bf The electron density profile} is shown in the lower panel of Fig.~\ref{A644}. The median profile of the simulated replica lies above that of the random cluster sample; however, its absolute values remain somewhat lower than the observed data. Notably, the mean density profile of the replica (indicated by the purple line) exhibits bumps in the central region that closely align with the observational measurements. This behavior may indicate the presence of substructures in the cluster core that are not fully aligned with the underlying gravitational potential, possibly reflecting recent merger activity, in agreement with observations of this cluster.
 
\end{itemize}

\subsection{Abell 1644} 

Abell 1644 (also A1644) is an SCC cluster that has been reported to exhibit clear signs of merging activity, evidenced by the presence of a double X-ray peak \citep[][]{2004ApJ...608..179R}. 

\begin{figure}
\centering
    \includegraphics[width=0.95\linewidth]{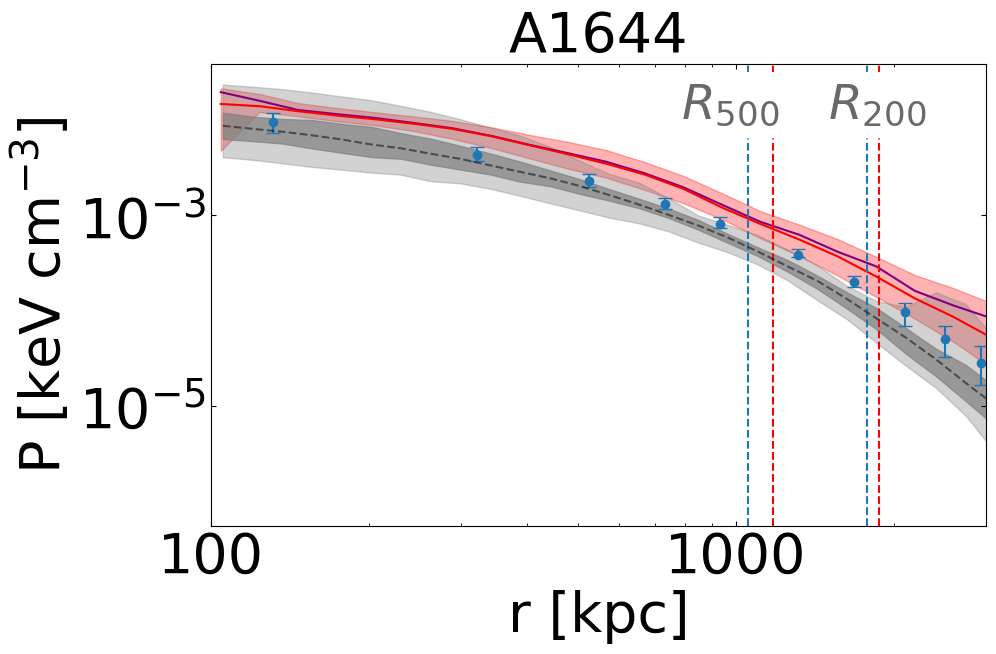}
    \includegraphics[width=0.95\linewidth]{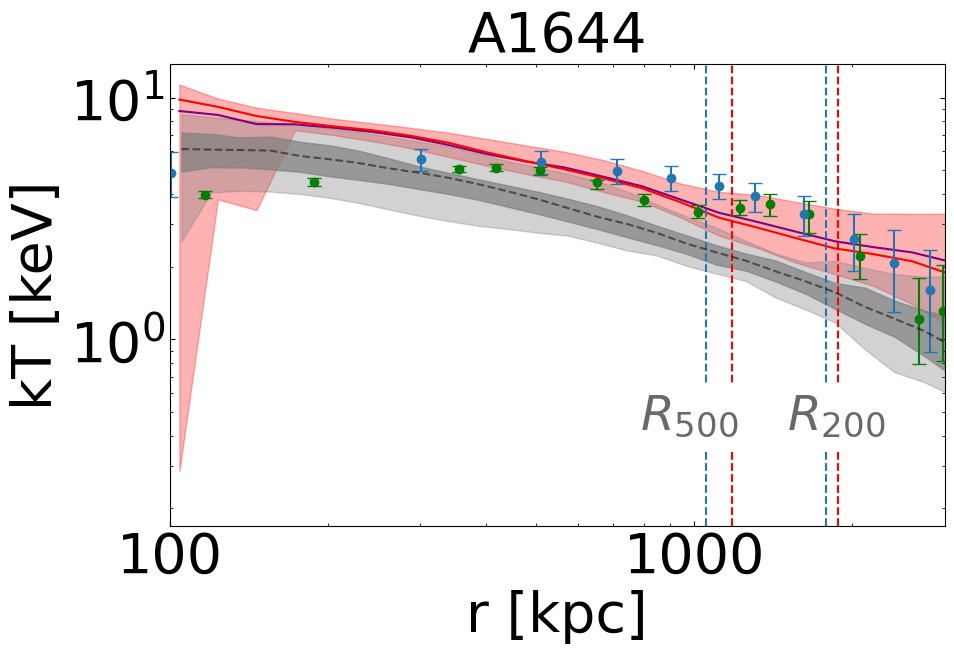}
    \includegraphics[width=0.95\linewidth]{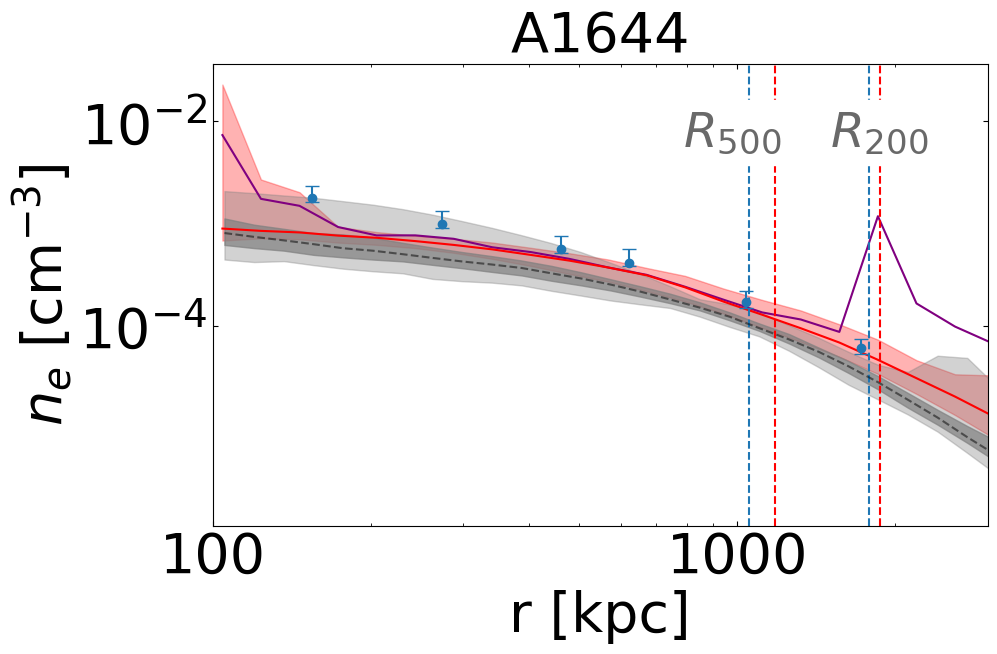}
    \caption{Same as Fig.~\ref{A644} but for the galaxy cluster Abell 1644. Observational data are from  \citet[][]{2019A&A...621A..39E}.}
    \label{A1644}
\end{figure}

\begin{itemize}
\item {\bf The pressure profile} of A1644 is shown in the upper panel of Fig.~\ref{A1644}. The simulated replica follows the overall trend of the observational data, lying near the lower edge of the 1$\sigma$ region across most radii. In terms of absolute values, the observed data points fall between those of the replica and the random cluster sample, suggesting that while the constrained simulation captures the general shape of the profile, it slightly overestimates the pressure across the radial range.

\item {\bf The temperature profile} (central panel) is well reproduced by the simulated replica from $r \sim 400$ kpc out to the cluster outskirts, with the profile lying clearly outside the 1$\sigma$ region of the random sample. This indicates that the constrained simulation captures the thermal structure of A1644 more accurately than typical simulated clusters of similar mass.

\item {\bf The electron density profile} of A1644 (lower panel) is nearly identical for the simulated replica and the random cluster sample across all radii. As a result, it is not possible to distinguish the replica from the random sample based on this profile alone.
\end{itemize}

\begin{figure}
\centering
    \includegraphics[width=0.95\linewidth]{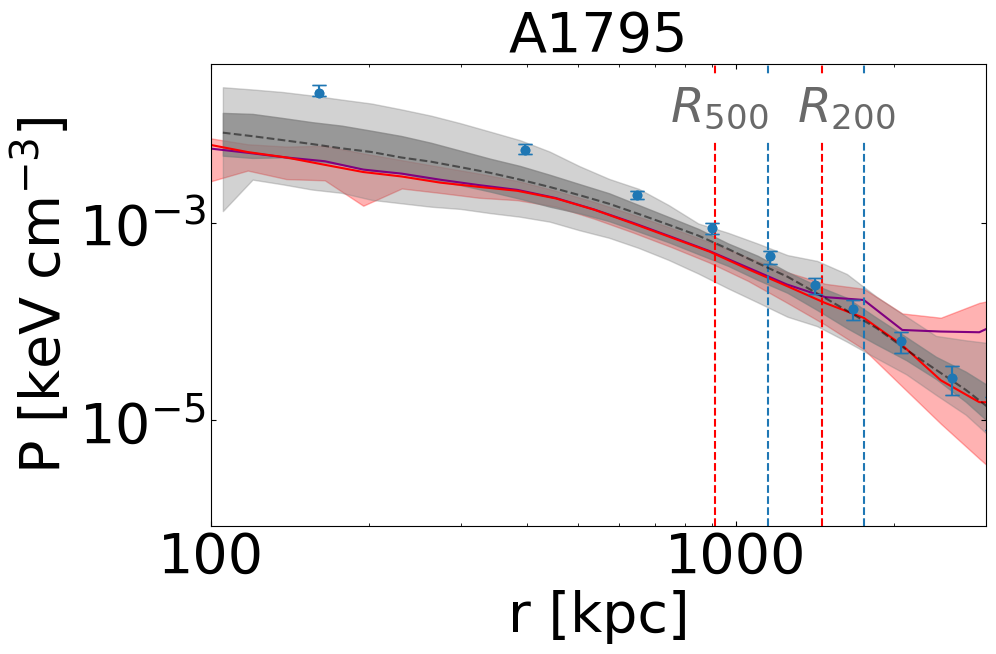}
    \includegraphics[width=0.95\linewidth]{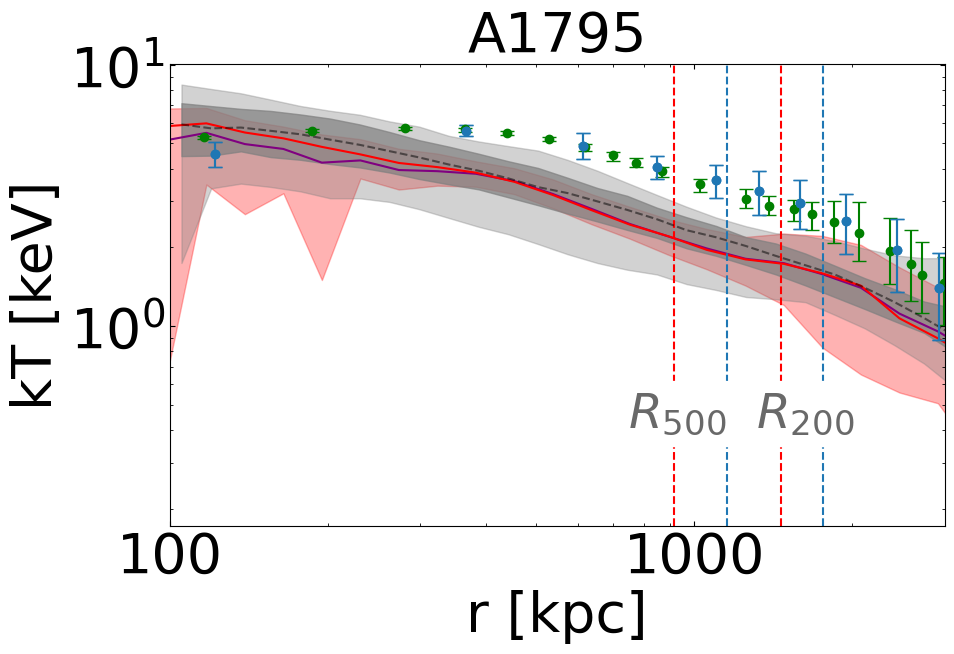}
    \includegraphics[width=0.95\linewidth]{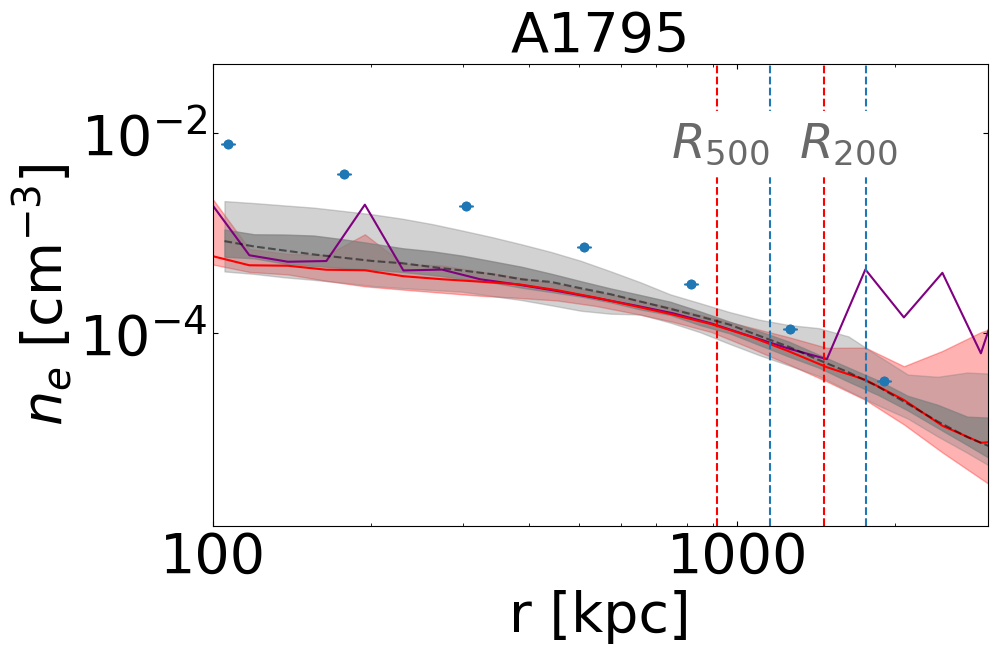}
    \caption{Same as Fig.~\ref{A644} but for the galaxy cluster Abell 1795. Observational data are from  \citet[][]{2019A&A...621A..39E}. 
    }
    \label{A1795}
\end{figure} 

\subsection{Abell 1795}
Abell 1795 (also A1795) is an SCC cluster that shows in its center a strong cooling, gas sloshing, and the presence of a cold front \citep[][]{2001ApJ...562L.153M}.

\begin{itemize}
\item {\bf The pressure profile} is shown in the upper panel of Fig.~\ref{A1795}. Within the central $\sim1000$ kpc, the observed pressure values are significantly higher than those recovered by either the simulated replica or the random sample. This offset likely reflects key features associated with strong cool-core (SCC) systems, such as enhanced cooling, gas sloshing, and the presence of a cold front \citep[][]{2001ApJ...562L.153M}. These processes tend to increase the central density and pressure, particularly when a cool gas filament and active feedback mechanisms are present, as reported in A1795 by \citet[][]{2001MNRAS.321L..20F}, \citet[][]{2001ApJ...560..187O}, and \citet[][]{2005MNRAS.361...17C}. At larger radii, beyond a few hundred kiloparsecs, the observed profile gradually converges toward the values predicted by the simulation.

\item {\bf The temperature profile} is shown in the central panel of Fig.~\ref{A1795}. The observational data (blue and green points) exhibit the characteristic decline in the innermost region, followed by a temperature peak at intermediate radii and a gradual decrease toward the outskirts—features indicative of a strong cool core surrounded by a hot intracluster medium (ICM). The simulated replica lies within the 1$\sigma$ region of the random sample but exhibits two mild bumps at $r \sim 300$ kpc and $r \sim 1100$ kpc, which qualitatively follow the observed structure.

However, the presence of gas sloshing \citep[][]{2001ApJ...562L.153M} and the prominent X-ray and H$\alpha$ filament observed within the central $\sim50$–$100$ kpc of A1795 \citep[][]{2005MNRAS.361...17C} likely introduce additional thermodynamic complexity that is not fully captured by the median simulated profile. These features, associated with multiphase gas and AGN feedback, may explain residual discrepancies in the core region.

\item {\bf The electron density profile} is shown in the lower panel of Fig.~\ref{A1795}. The observed profile exhibits a sharp central peak, characteristic of a SCC system. The simulated replica fails to reproduce both the high central electron densities and the steep inner slope, likely due to resolution limitations that hinder accurate modeling of the dense core region.

Even more, the estimated values for  \( r_{500}\) and  \( r_{200}\) are notably lower in the simulation than the ones estimated from the observations. Thus, higher resolution and possibly a better treatment of cooling and feedback would be necessary to reproduce the ICM properties of A1795.

\end{itemize}

\begin{figure}
\centering
    \includegraphics[width=0.95\linewidth]{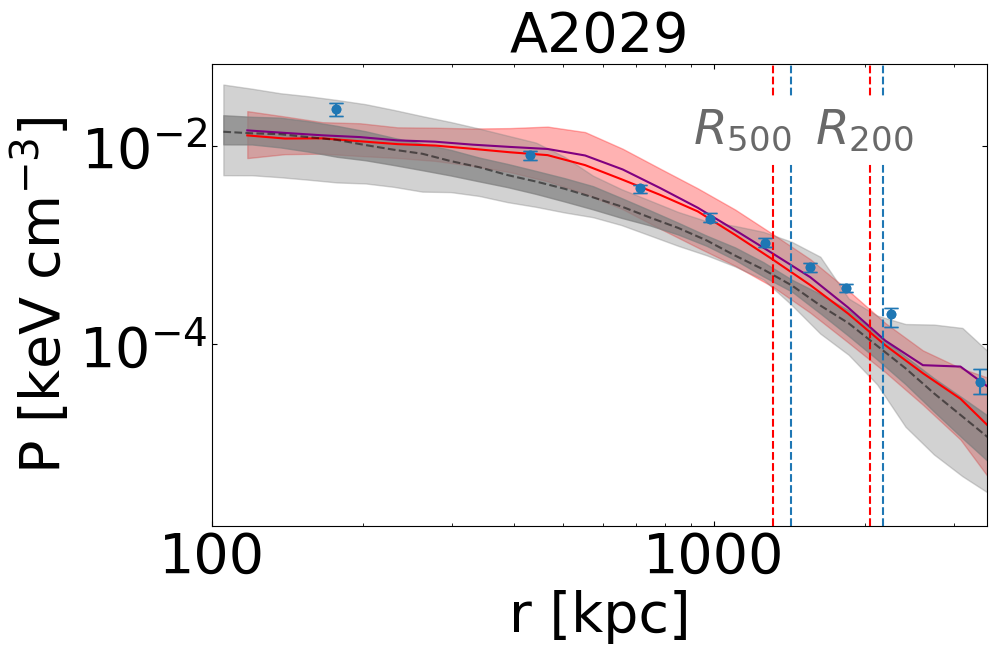}
    \includegraphics[width=0.95\linewidth]{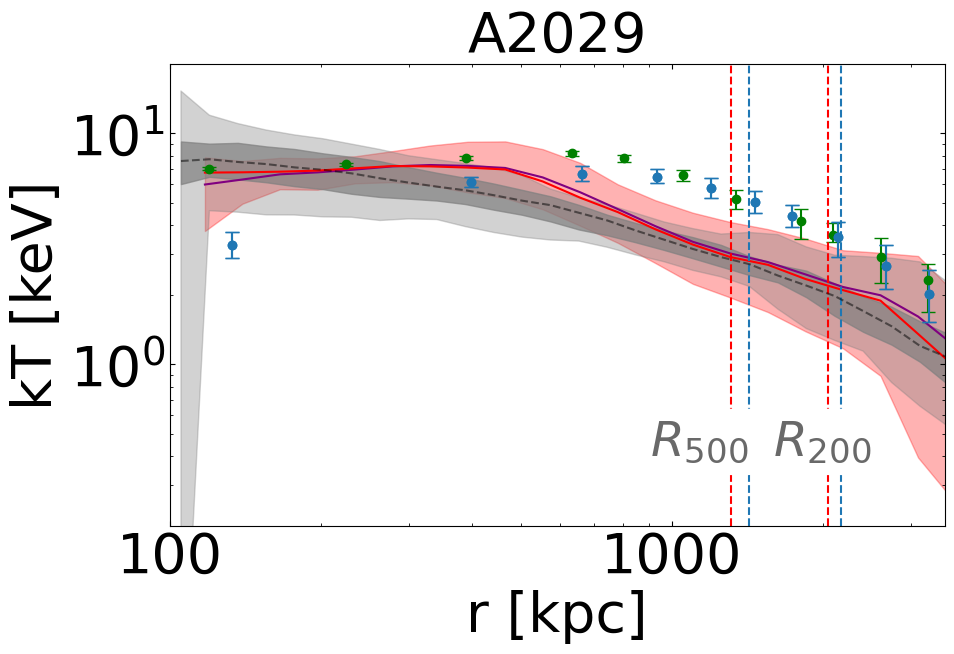}
    \includegraphics[width=0.95\linewidth]{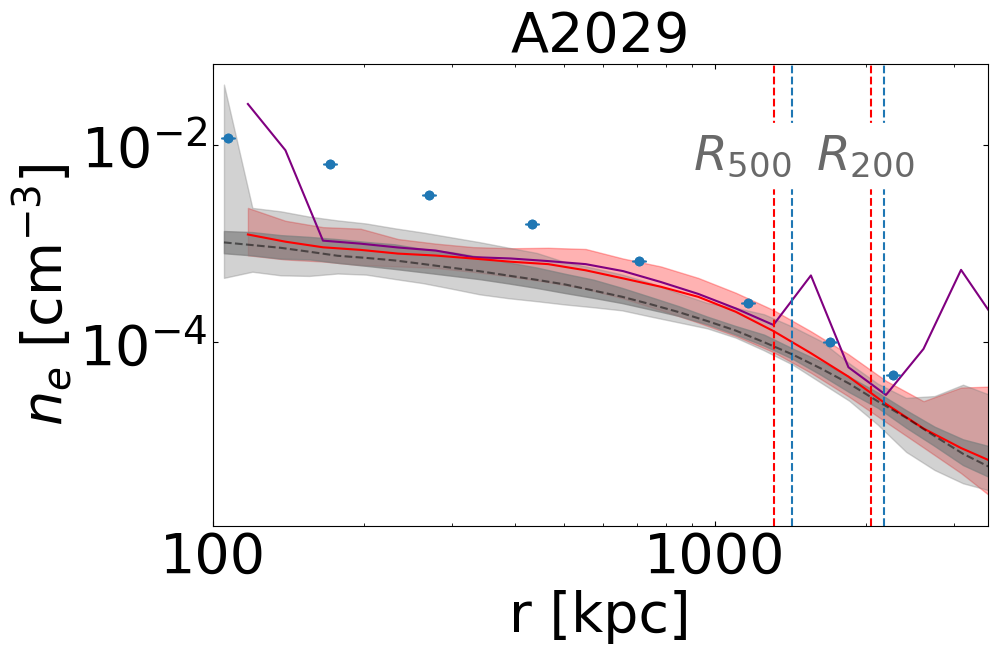}
    \caption{Same as Fig.~\ref{A644} but for the galaxy cluster Abell 2029. Observational data are from  \citet[][]{2019A&A...621A..39E}.}
    \label{A2029}
\end{figure}

\begin{figure}
\centering
    \includegraphics[width=0.95\linewidth]{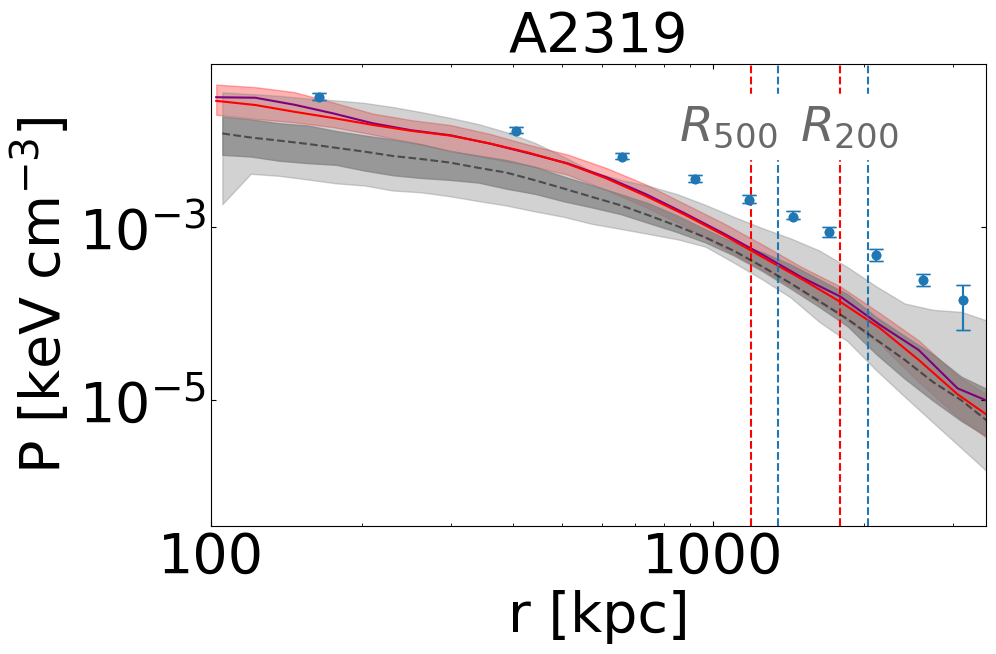}
    \includegraphics[width=0.95\linewidth]{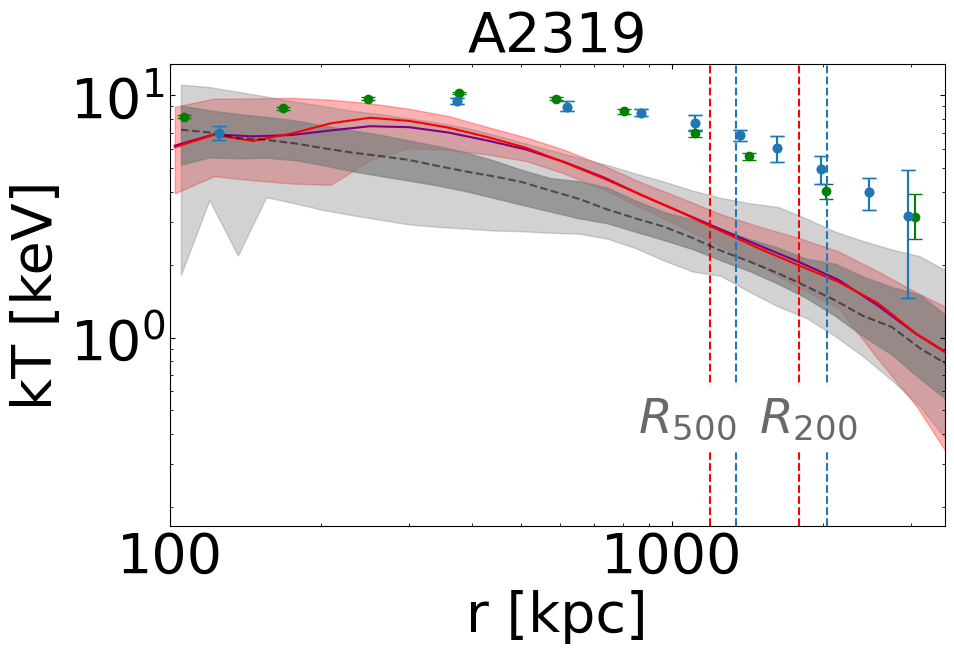}
    \includegraphics[width=0.95\linewidth]{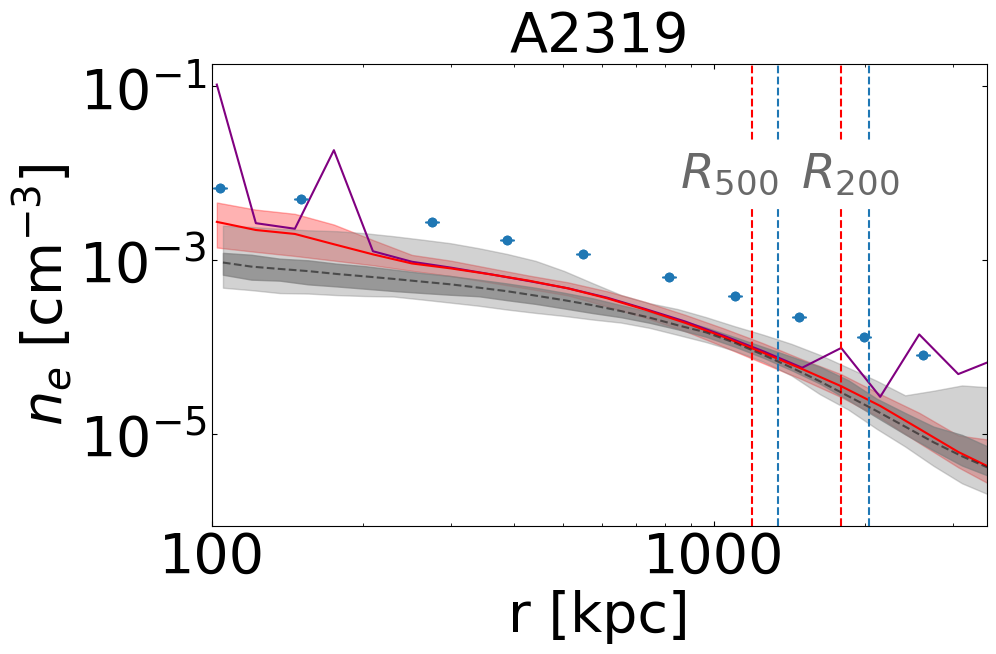}
    \caption{Same as Fig.~\ref{A644} but for the galaxy cluster Abell 2319. Observational data are from  \citet[][]{2019A&A...621A..39E}. 
    }
    \label{A2319}
\end{figure}

\subsection{Abell A2029}

Abell 2029 (also A2029) is a well-studied SCC cluster that features a cold front, as identified by \citet[][]{2003ASPC..301...37M}. The core has been extensively analyzed with {\it Chandra}, including a detailed study by \citet[][]{2004rcfg.proc...19C}. Thermodynamic and chemical properties reveal a dynamic and well-regulated environment.  Figure \ref{A2029} presents the deprojected thermodynamic profiles of A2029 from XCOP observations compared to simulated profiles. 

\begin{itemize}
\item {\bf The pressure profile} is shown in the upper panel of Fig.~\ref{A2029}. The observed data begin at values of a few $10^{-2}\,\text{keV\,cm}^{-3}$ in the core ($ r \sim 100$–$200 $ kpc) and decline smoothly by approximately two to three orders of magnitude out to $r \sim 1$ Mpc. The median pressure profile of the simulated A2029 replica follows the observational trend closely, with its 1$\sigma$ region encompassing the observed data across most of the radial range. In contrast, the random cluster sample fails to capture the observed profile, with the data consistently lying outside its 1$\sigma$ region, and thus highlighting the improved performance of the constrained simulation in reproducing the thermodynamic structure of A2029.

\item {\bf The temperature profile}, shown in the middle panel of Fig.~\ref{A2029}, indicates that the deprojected observational values are relatively high in the core ($ \sim 9$–$10\,\mathrm{keV} $), rising to a broad peak of $ \sim 11$–$13\,\mathrm{keV} $ at a few hundred kiloparsecs, before declining beyond $r \sim 500$–$700 $ kpc. The median temperature profile of the simulated A2029 replica agrees well with the overall shape of the observed profile (green and blue points). While the simulation peak occurs slightly closer to the center ($\sim 500$ kpc) than the observed peak at $\sim 600$ kpc, both exhibit consistent thermal structure.

This behavior lies entirely outside the 1$\sigma$ region of the random cluster sample, indicating that only the constrained replica is able to reproduce the observed temperature distribution. Beyond the peak, both the simulation and observations show a declining trend, with the simulated profile exhibiting a slightly steeper drop. Overall, this agreement suggests that the simulation captures the thermodynamic structure of A2029’s hot intracluster medium with a high degree of fidelity.

\item {\bf The electron density profile} is shown in the lower panel of Fig.~\ref{A2029}. As is typical for massive clusters, A2029 exhibits a sharply peaked central electron density, reaching values of $n_e \sim 3 \times 10^{-2}\,\mathrm{cm}^{-3}$ at $r \sim 100$ kpc, followed by a steep decline over several orders of magnitude with increasing radius. This steep central profile presents a challenge for the simulation, which underestimates the electron density within the inner $\sim700$ kpc, likely due to the resolution limitation. Beyond this radius, the simulated and observed profiles begin to converge, suggesting that the large-scale gas distribution is reasonably well captured.
\end{itemize}

\subsection{Abell 2319}

Abell 2319 (also A2319) is a well-studied merger cluster, known for its large core region extending beyond 100 kpc and a prominent cold front \citep[][]{1997NewA....2..501F, 1999ApJ...525L..73M, 2004ApJ...604..604O}.

\begin{itemize}
\item {\bf The pressure profile} of A2319 is shown in the upper panel of Fig.~\ref{A2319}. The observed deprojected pressure profile (blue points) declines steadily from $\sim10^{-2}\,\mathrm{keV\,cm}^{-3}$ in the core to $\sim10^{-5}\,\mathrm{keV\,cm}^{-3}$ beyond $r_{500}$. The median profile of the simulated replica (red line) tracks the observations closely in the inner regions, lying outside the 1$\sigma$ region of the random cluster sample. However, at intermediate and large radii, the simulation begins to systematically underestimate the pressure.

This deviation likely reflects limitations in the simulation's ability to capture the full extent of gas compression and redistribution associated with the cluster's complex merger history. Additionally, the simulated estimates for $r_{500}$ and $r_{200}$ are lower than the observationally inferred values, consistent with the underestimation of thermal pressure in the outer regions.

\item  {\bf The temperature profile} is shown in the middle panel of Fig.~\ref{A2319}. The observed data exhibit a broad peak of $ \sim10$–$12\,\mathrm{keV} $ around $r \sim 400$ kpc. The simulated replica shows a similar peak, though slightly offset inward at $r \sim 300$ kpc, and lies outside the 1$\sigma$ region of the random sample, indicating a better match to the observations. At larger radii, the temperature profile of the simulated replica declines more steeply than the observed data, although it still broadly follows the overall trend. This suggests that the constrained simulation captures the thermal structure of A2319 reasonably well, but may slightly underestimate the temperature in the outskirts.

\item {\bf The electron density profile}, shown in the lower panel of Fig.~\ref{A2319}, exhibits a steep decline with radius, with the simulated replica and the random cluster sample following a slope similar to that of the observations. However, both systematically underestimate the observed electron densities by a nearly constant factor of $\sim3$ across all radii. This suggests a persistent bias in the simulation, possibly due to limitations in the modeling of gas physics or resolution limitations.

\end{itemize}

Generally, the simulation reproduces the general trends of the thermodynamic profiles of A2319 but fails to replicate their absolute values, indicating a possible systematic bias. In the core, these biases are likely driven by an insufficient representation of gas compression and cold front dynamics, while in the outskirts, they could stem from an underestimation of shock heating and turbulence. The smoothing of substructures and the assumption of quasi-hydrostatic equilibrium in the simulations contribute to these discrepancies, as well as the underestimation of r200 and r500, particularly in systems like A2319 that are far from equilibrium.\\

\subsection{Abell 3158}

\begin{figure}
\centering
    \includegraphics[width=0.95\linewidth]{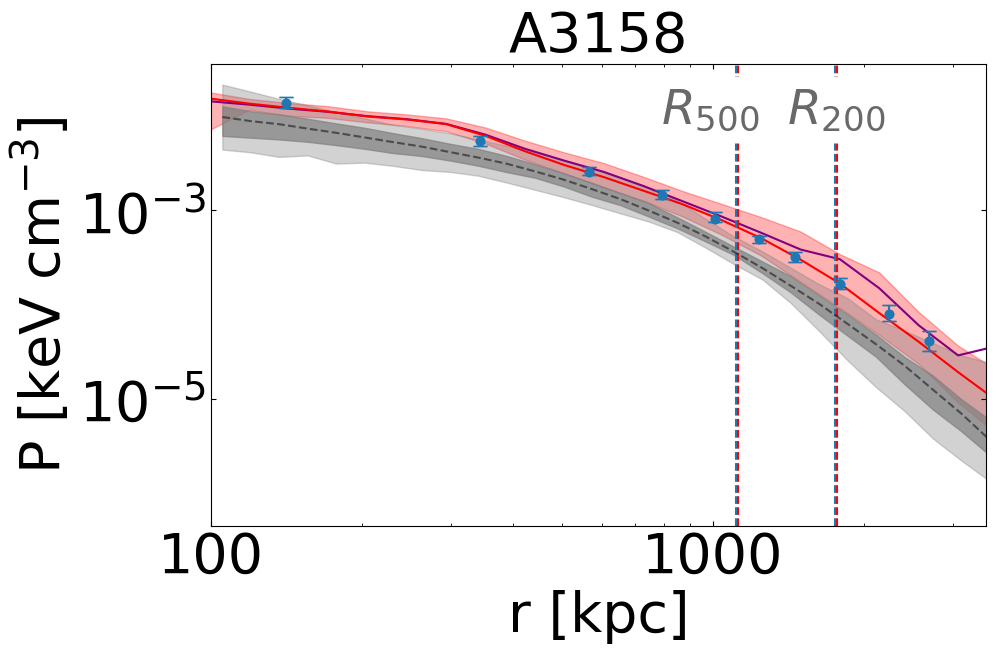}
    \includegraphics[width=0.95\linewidth]{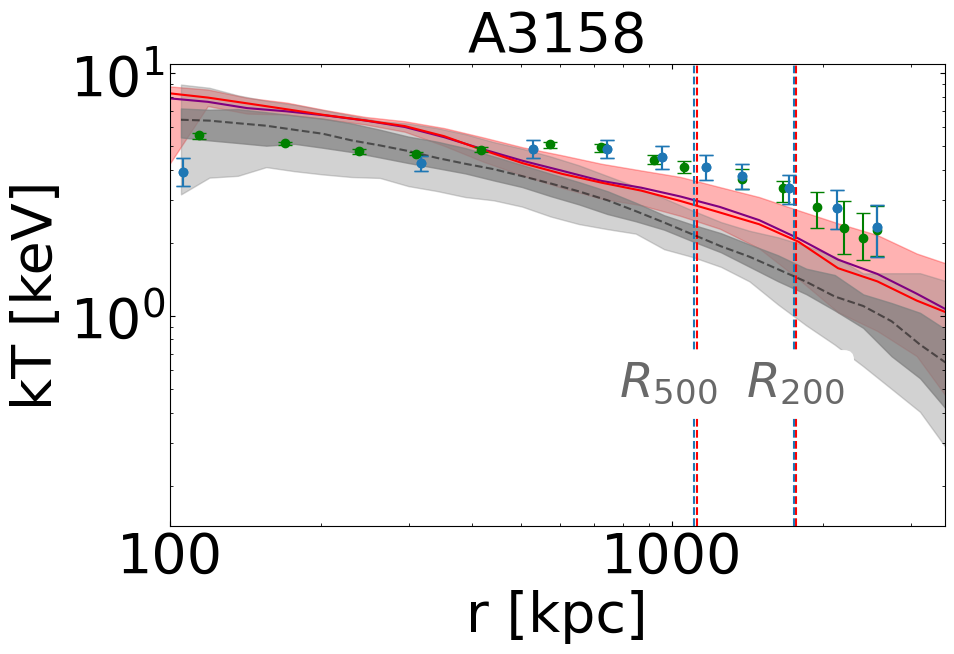}
    \includegraphics[width=0.95\linewidth]{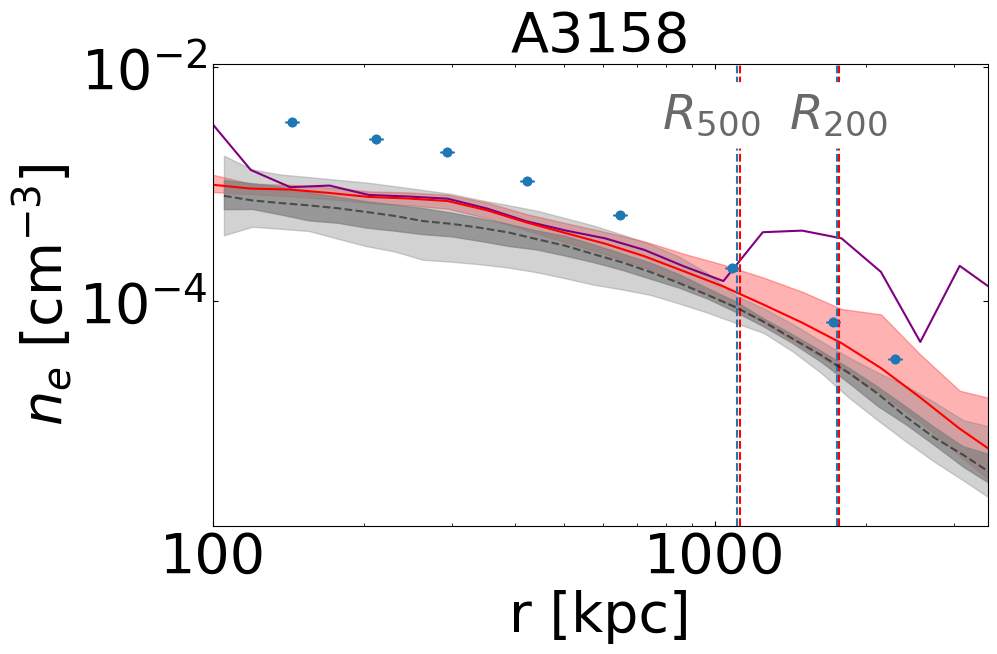}
    \caption{Same as Fig.~\ref{A644} but for the galaxy cluster Abell 3158. Observational data are from  \citet[][]{2019A&A...621A..39E}.}
    \label{A3158}
\end{figure}

Abell 3158 (also A3158) is classified as a relaxed cluster based on the velocity dispersion of its member galaxies \citep[][]{2006MNRAS.367.1463L}. The cluster’s X-ray emission is elliptical in shape, and it hosts two central cD galaxies, with one located at the X-ray peak. Despite its relaxed classification, the cluster does not have a bright core, exhibiting a relatively low central electron density of approximately \(5 \times 10^{-3} \, \mathrm{cm^{-3}}\). These characteristics, coupled with its thermodynamic properties, provide a framework to interpret the patterns observed in its pressure, temperature, and density profiles.

\begin{itemize}
\item {\bf The pressure profile} is shown in the upper panel of Fig.~\ref{A3158}. The simulation reproduces the observed pressure profile remarkably well, with the median of the A3158 simulated replica closely following the observational data points across most radii. Notably, the profile lies completely outside the region of the random cluster sample, underscoring the distinctiveness and effectiveness of the constrained simulation in capturing the thermodynamic structure of A3158. \\

\item {\bf The temperature profile} is presented in the middle panel. The simulation performs well in terms of absolute values, although the simulated replica shows a slowly decreasing trend with radius, while observations show a flatter profile. However, the core and intermediate regions (\(r < r_{500}\)) successfully reproduce the observed peak temperature of approximately 5.7 keV in the center. For r>500 kpc the values of observations and the replica lie outside the random sample region, showing a correspondence between the simulated replica and observations difficult to explain by random chance. \\

is presented in the middle panel of Fig.~\ref{A3158}. The simulation performs well in terms of absolute temperature values, although for $r<500$ kpc, the simulated replica exhibits a mildly declining trend with radius, while the observations show a comparatively flatter profile. At intermediate radii ($r \sim 500$ kpc), the replica successfully reproduces the observed central temperature peak of approximately 5.7 keV. Beyond $r > 500$ kpc, both the observed and simulated profiles lie outside the full range covered by the random cluster sample, highlighting a level of agreement between the replica and observations that is unlikely to arise from random chance. \\

\item {\bf The electron density profile} from observations shown in the lower panel shows a rather low density in the center; nevertheless, due to resolution limitations this simulation is not capable of achieving central values much higher than $10^{-3}$, falling short in reproducing the inner electron densities of the cluster cores. Observations and simulations start converging again at \(\sim r_{500}\). \\

\end{itemize}

\begin{figure}
\centering
    \includegraphics[width=0.95\linewidth]{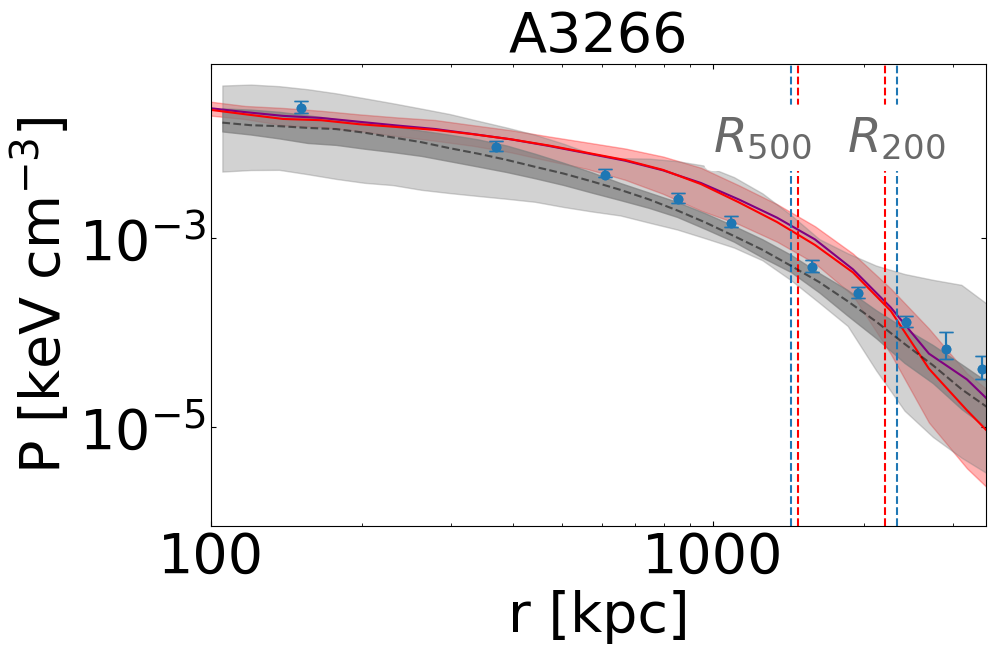}
    \includegraphics[width=0.95\linewidth]{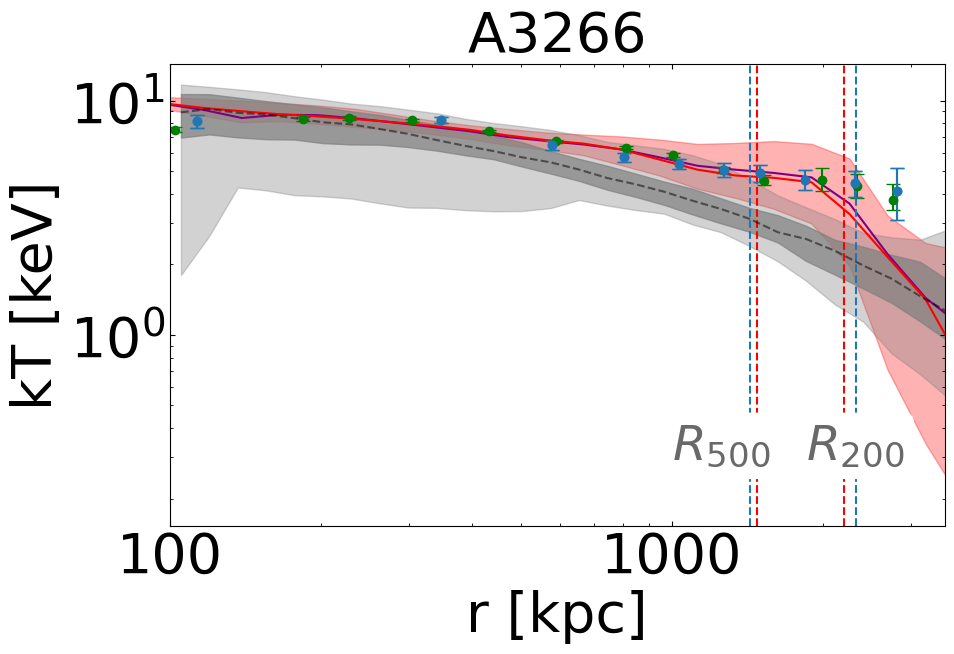}
    \includegraphics[width=0.95\linewidth]{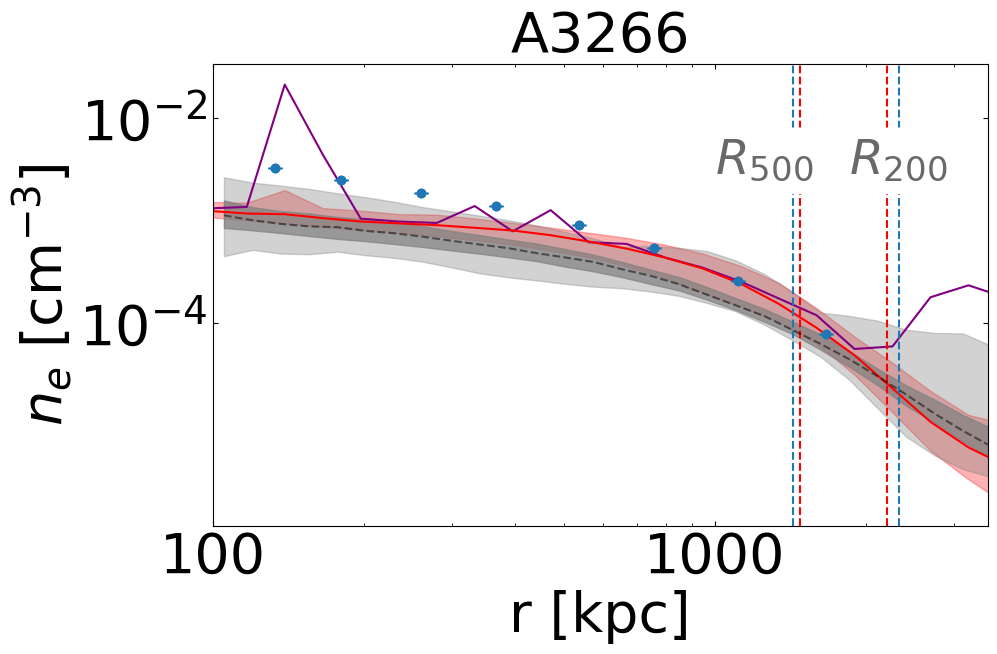}
    \caption{Same as Fig.~\ref{A644} but for the galaxy cluster Abell 3266. Observational data are from  \citet[][]{2019A&A...621A..39E}.}
    \label{A3266}
\end{figure}
\begin{figure}[ht!]
\centering
    \includegraphics[width=0.95\linewidth]{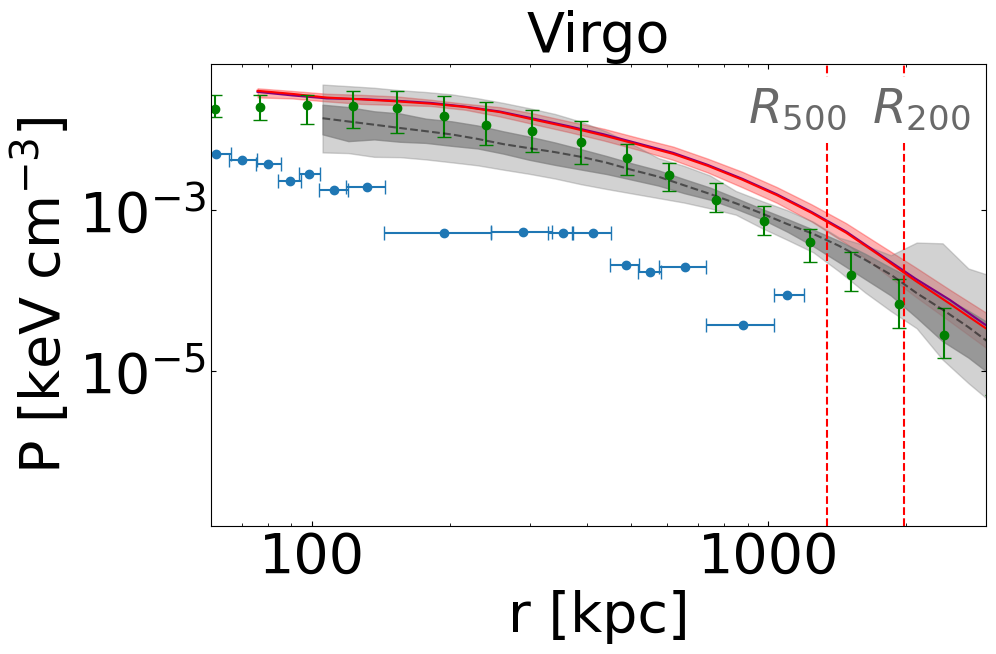}
    \includegraphics[width=0.95\linewidth]{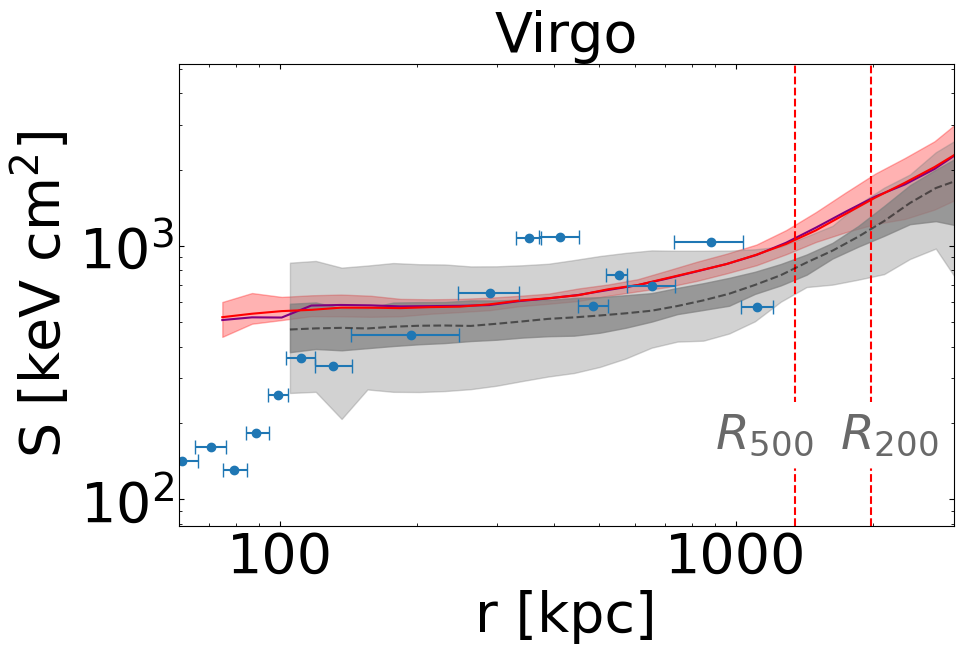}
    \includegraphics[width=0.95\linewidth]{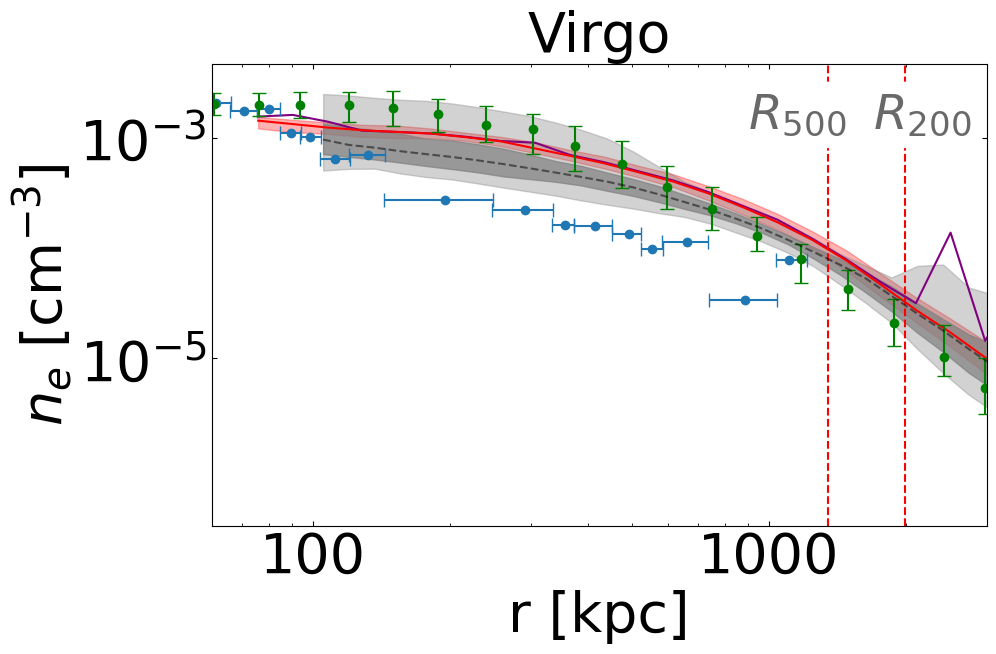}
    \caption{Deprojected thermodynamic profiles of the Virgo cluster, comparing simulation results to observational data. The panels, from left to right, show the pressure, entropy, and electron density profiles. The red line represents the median profile of the simulated Virgo cluster, while the red shaded region indicates the 1 
    $\sigma$ scatter of particles within each spherical-shell bin. The purple line represents the mean profile of the simulation.  The gray line represents the median profile of the random cluster sample; the dark gray shaded region denotes the corresponding $1\sigma$ spread, and the light gray region shows the full range covered by the sample at each radial bin. Observational data from \citet[][]{2011MNRAS.414.2101U} are shown as blue points. The green points represent 3D profiles from a Virgo replica run using the RAMSES code presented by \citet[][]{2024A&A...682A.157L}, which employs different subgrid physics and resolution (dark matter particle mass of $3\times10^7$ $M_{\odot}$). The vertical dashed lines indicate $r_{500}$ and $r_{200}$ from the simulated replica. 
    }
    \label{Virgo}
\end{figure}

\subsection{Abell 3266}

Abell 3266 (also A3266) is a well-known merging cluster \citep[see e.g.][]{2002ApJ...577..701H}.

\begin{itemize}
\item {\bf The pressure profile}, shown in the upper panel of Fig.~\ref{A3266}, demonstrates good agreement between the simulated replica and the observations in the core and intermediate regions ($r < r_{500}$), suggesting that the simulation captures the broad thermodynamic structure of the cluster reasonably well. However, we also note that the random cluster sample reproduces the observational data comparably well in this case, indicating that the pressure profile of A3266 is not uniquely matched by the constrained simulation.

\item {\bf The temperature profile}, shown in the middle panel of Fig.~\ref{A3266}, reflects the cluster's disturbed thermodynamic state, exhibiting a relatively flat temperature distribution in the core and intermediate regions. This behavior is consistent with turbulent mixing and merger-induced heating, which act to suppress the development of a steep temperature gradient. Notably, this feature is not reproduced by the random cluster sample, underscoring the distinctiveness of the A3266 replica in capturing the observed thermal structure.

\item {\bf The electron density}, shown in the lower panel of Fig.~\ref{A3266}, provides further evidence of merger-induced effects in A3266. While the simulation matches the observed density reasonably well in the intermediate and outer regions, it systematically underestimates the electron density in the inner regions ($r < 400$ kpc), as expected given the resolution limits of the simulation. Interestingly, a noticeable deviation between the mean and median profiles occurs near $r_{500}$ for the simulated replicq, suggesting the presence of substructure—likely associated with an ongoing merger—that disrupts the symmetry and smoothness of the gas distribution.
 \\

\end{itemize}

\begin{figure*}
  \centering
    \includegraphics[width=0.95\linewidth]{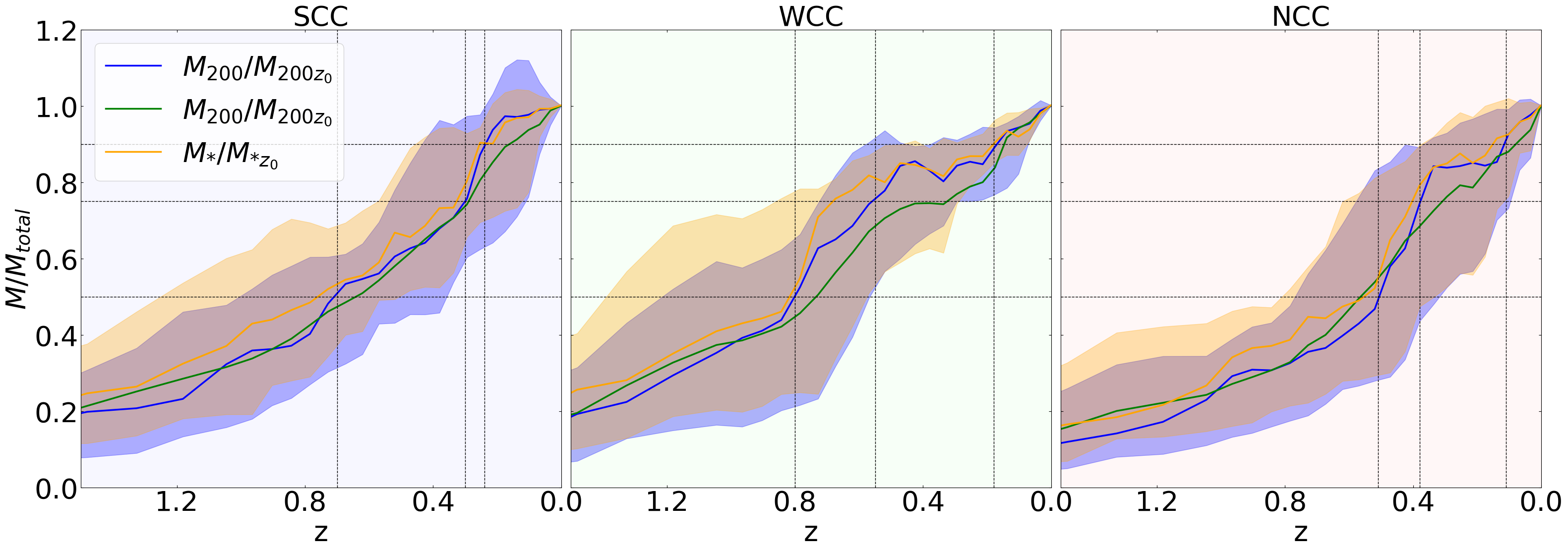}
    \caption{Evolution of the mass assembly history for clusters classified as cool-core, weak cool-core, and non-cool-core based on observational criteria. The panels show the total median and mean mass growth as well as the total stellar mass growth, normalized to the final mass, as a function of redshift. The blue line represents the median \( M_{200c}/M_{200, z=0} \) mass growth, while the green line shows the mean \( M_{200c}/M_{200, z=0} \) mass growth. The orange line represents the median stellar mass growth, \( M_{\star}/M_{\star, z=0} \). The shaded regions indicate the 1\( \sigma \) scatter for the total median mass and the median stellar mass for each classification. The vertical dashed lines mark the redshifts at which 50\%, 75\%, and 90\% of the total mass was assembled.}
    \label{mass history median}
\end{figure*}  

\subsection{The Virgo  Cluster}

The Virgo Cluster is the nearest massive galaxy cluster, located at a distance of 16 Mpc \citep[][]{2007ApJ...655..144M, 2012ApJS..200....4F}. Unlike many other well-studied clusters, Virgo is a dynamically young system, exhibiting significant substructure and signs of a late-time merger and ongoing formation \citep[e.g.,][]{2014A&A...570A..69B, 2016MNRAS.460.2015S, 2021MNRAS.504.2998S}. X-ray studies, such as the one performed by \citet[][]{2011MNRAS.414.2101U} have revealed that the cluster has a relatively shallow density profile, with a power-law index of $\beta$ = 1.21 $\pm$ 0.12, which is lower than what is typically observed in more massive, relaxed clusters. Additionally, the entropy profile follows a gravitational collapse-like power law (\( K \propto r^{1.1} \)) within 450 kpc, but beyond this radius, it flattens significantly, falling below theoretical expectations by a factor of 2–2.5 \citep[][]{ 2002A&A...391..841E, 2011MNRAS.414.2101U}. This behavior has been attributed to clumping in the ICM, which artificially enhances the observed density while reducing the inferred entropy \citep[][]{2011Sci...331.1576S, 2011ApJ...731L..10N}.  

Due to its non-relaxed state and complex morphology, the results of the de-projection analysis, which assumes spherical symmetry, should be interpreted with caution \citep[][]{2004astro.ph.12536M}. In fact,  \citet[][]{2011MNRAS.414.2101U} noted that systematic uncertainties due to deviations from spherical symmetry are significantly larger than measurement errors, complicating direct comparisons with theoretical models or simulations.  

Fig.~\ref{Virgo} presents the deprojected thermodynamic profiles of the Virgo Cluster from \citet[][]{2011MNRAS.414.2101U} alongside the simulated profiles. The vertical dashed lines mark \( r_{500} \) and \( r_{200} \). Additionally, green points represent the deprojected profiles of a Virgo replica simulated with the RAMSES code as presented in \citet[][]{2024A&A...682A.157L}, which employs a different subgrid physics model and resolution (DM mass of \(10^9\) \(M_\odot\) in the RAMSES replica, while $2.9 \times10^9$ \(M_\odot\) in the SLOW replica). Notably, both simulations (the main one and RAMSES) agree with each other, despite these differences in the implementation of code and subgrid physics. 

\begin{itemize}
\item   {\bf The pressure profile } (upper panel of Figure \ref{Virgo}) shows notable discrepancies between simulations and observations, where observed data points systematically fall below the simulation. This may be due to departure from hydrostatic equilibrium, as turbulence and bulk motions in the Virgo ICM can redistribute pressure, leading to deviations from hydrostatic models. Additionally, projection effects in the observational de-projection analysis may contribute to the observed offset. The agreement between the two simulations suggests that this discrepancy is not simply a result of the simulation's subgrid physics or resolution. Instead, it may arise from a combination of observational systematics, Virgo’s non-relaxed state, and potential limitations in the simulation model of ICM turbulence and heating.  

\item   {\bf The entropy profile }  (middle panel of Figure \ref{Virgo}) shows a significant mismatch, with observed entropy values well below the simulated predictions. The simulated profile predicts a more gradual increase, while observations indicate a steeper rise followed by a flattening. This suggests that the simulation may overpredict heating processes or underestimate gas mixing, leading to systematically higher entropy values. Given the cluster’s unrelaxed state, deviations from spherical symmetry may introduce further biases in the deprojected entropy values. The fact that the RAMSES simulation reproduces the same trends reinforces the robustness of the simulated entropy structure, regardless of the numerical implementation or resolution. However, the discrepancies could also indicate that the simulations fail to fully capture some of the physical processes shaping the entropy profile, such as turbulence, conduction, or small-scale gas clumping.

\item {\bf The electron density profile} (lower panel of Figure \ref{Virgo}) also exhibits systematic differences. While the general decline in density is captured, the simulated cluster maintains a higher central density than the one observed. Observations indicate a more diffuse gas distribution, which could result from large-scale gas motions, substructure, or incomplete virialization. The agreement between the two simulations suggests that these discrepancies are not only an artifact of the simulation technique, but a consequence of a combination of observational de-projection assumptions, departures from hydrostatic equilibrium, and possibly limitations in how the simulations model the core structure of the Virgo cluster.  

\end{itemize}
Overall, the Virgo Cluster presents significant challenges for direct comparisons between simulations and observations due to its dynamically non-relaxed nature. The large uncertainties in the de-projection analysis, as highlighted by \citet[][]{2011MNRAS.414.2101U}, suggest that some observed-simulated discrepancies may stem from the limitations of assuming spherical symmetry and hydrostatic equilibrium. However, the possibility remains that current simulations may not fully capture some aspects of Virgo’s complex thermodynamic state, such as non-thermal pressure support, gas clumping, or turbulent mixing. Future work could explore non-spherical modeling approaches, alternative subgrid physics models, or dynamical state indicators to better characterize the Virgo cluster’s complex ICM.

\subsection{Successes and Limitations in Reproducing Local Clusters} 

The comparison between simulated replicas and observed thermodynamic profiles reveals that our constrained simulations reproduce several local clusters with remarkable fidelity, while others remain more challenging. Clusters such as Perseus, Coma, A85, A119, A1644, A2029, A3158, and A3266 show excellent agreement across pressure, temperature, and, in some cases, entropy profiles, often lying well outside the region covered by randomly selected clusters of similar mass. This indicates that the constrained replicas capture key features of the clusters’ thermodynamic structure, even for systems spanning a wide range of dynamical states and core properties, that are not reproduced by the random sample.

On the other hand, clusters such as A644, A1795, A2319, and Virgo highlight important limitations. In these cases, discrepancies emerge in the core or outskirts, most probably tied to the inability to fully model disturbed ICM states. These mismatches point to areas for improvement: enhanced resolution, more sophisticated AGN feedback models, and better treatment of turbulence and non-thermal pressure support.

Rather than being shortcomings, these tensions offer valuable diagnostic power. By identifying which clusters are more or less faithfully reproduced, we gain insight into which physical processes are captured reliably by the simulation and which require further refinement. These one-to-one comparisons thus provide a promising path toward improving our understanding of ICM physics and increasing the predictive power of constrained simulations of the local Universe.

\section{Connecting formation histories with late time cluster core states}
\label{subsec:cores}
The assembly history of a galaxy cluster plays a crucial role in determining its present-day thermodynamic state \citep[][]{2005MNRAS.364..909V,2008ApJ...675.1125B,2012ARA&A..50..353K,2015ApJ...806...68L}. Current simulations, including the SLOW suite, face challenges in accurately reproducing the thermodynamic profiles of the innermost regions of galaxy clusters. In particular, they struggle to replicate the observed differences in central slopes between SCC, weakly cool-core (WCC), and NCC clusters \citep[see e.g.][]{2015ApJ...813L..17R, 2017MNRAS.467.3827P, 2018MNRAS.481.1809B, 2025A&A...694A.232G}. These discrepancies likely arise from an incomplete understanding of subgrid physics, particularly regarding cooling, turbulence, and feedback mechanisms. However, a key strength of SLOW constrained simulations is their ability to faithfully reproduce the large-scale structures of the local Universe, ensuring that the formation histories of individual clusters are well captured \citep[see, e.g.,][]{2016MNRAS.460.2015S}. With our simulations successfully matching the observed thermodynamic profiles at intermediate and large radii—where our lack of resolution and subgrid physics does not introduce significant biases—we can be confident that the formation pathways of these structures are realistically constrained. This enables us to investigate the connection between a cluster’s assembly history and its present-day classification as CC, WCC, or NCC. The classification of clusters follows the criteria established in the HIGFLUGCS catalog \citep[][]{2010A&A...513A..37H}.

Fig.~\ref{mass history median} illustrates the median evolution of total mass (\( M_{200} \)) and stellar mass (\( M_*\)), normalized to their respective values at \( z=0 \), for the clusters in our simulated volume classified by observations as CC, WCC, and NCC clusters respectively. A clear trend emerges: current-day CC clusters tend to finish their mass assembly earlier than their WCC and NCC counterparts. Current day CC have assembled a larger fraction of their total mass already at $z\approx0.25$ while their WCC and NCC counterparts still grow in mass significantly after $z\approx0.25$. The vertical markers in Figure \ref{mass history median} indicate the redshifts at which 50$\%$, 75$\%$, and 90$\%$ of the total mass has been accreted.
Clearly visible for CC clusters is an extended plateau from z $\approx$ 0.25 to z $\approx$ 0.  This suggests a more rapid formation history before, followed by a late-time relaxation. In contrast, NCC clusters accumulate their mass more gradually, and a significant fraction of their growth occurs more abruptly at recent times, indicating fast accretion episodes through significant mergers. WCC clusters exhibit an intermediate accretion pattern, bridging the gap between CC and NCC populations. Their mass assembly shows subsequent increases from z$\approx$1.0 onward, but these are less abrupt compared to the NCC case. This increased merger rate of NCC clusters is in line with recent findings based on the driver of the transition between CC and NCC clusters based on the {\it Magneticum} simulations \citep{2025A&A...694A.232G}. 

This distinction in mass growth histories is also reflected in the evolution of the stellar mass component. The stellar mass growth curves (yellow lines in Figure \ref{mass history median}) reveal that CC clusters build up their stellar content earlier and more efficiently compared to NCC clusters. This early stellar mass accumulation is consistent with a formation scenario where CC clusters experience early, intense cooling and subsequent star formation within their cores. Conversely, NCC clusters show a delayed stellar mass growth, likely due to a more prolonged merger history that disrupts early cooling and extends the timescales for stellar assembly.

The delayed assembly of NCC clusters aligns with the expected more disturbed dynamical states observed at \( z=0 \). Mergers inject turbulence into the ICM, suppressing the formation of a central cool core by redistributing entropy and mixing high- and low-entropy gas. This mechanism naturally explains why NCC clusters exhibit flatter entropy profiles and weaker central cooling signatures compared to CC clusters. The enhanced scatter in the total mass accretion histories of NCC clusters further supports the idea that these systems undergo more stochastic growth, with frequent late-time mergers playing a dominant role in shaping their final structure.

WCC clusters represent an intermediate evolutionary state, exhibiting characteristics of both CC and NCC populations. Their accretion histories suggest a mix of early mass assembly with later minor mergers that do not fully disrupt their cores. This transitional nature may indicate that WCC clusters evolve into NCC clusters over time if additional late-time mergers continue to inject energy into the ICM.

Overall, the connection between merger history and cluster classification at \( z=0 \) is evident in both the total and stellar mass assembly histories. CC clusters, which form earlier and more quiescently, retain their dense, cooling cores, while NCC clusters, shaped by extended merger activity, remain dynamically disturbed with higher entropy cores. The SLOW simulations successfully reproduce these trends, offering a detailed view of how individual formation pathways influence the present-day structure of local galaxy clusters.

\section{Conclusions}
\label{sec:Conclusions}

In this study, we examined the thermodynamic properties of local galaxy clusters using the constrained cosmological simulation AGN1536$^3$ from the SLOW project. By performing one-to-one comparisons with observed systems, we assessed the ability of the simulation to reproduce late-time profiles in the Local Universe and explored the physical processes shaping them. The close agreement between simulated and observed deprojected profiles suggests that constrained simulations effectively capture the large-scale formation histories of individual clusters, providing a reliable framework for connecting past accretion and merger events to present-day thermodynamic states.  

Our findings highlight several key aspects of cluster evolution:  

\begin{enumerate}

\item The SLOW simulations successfully reproduce the thermodynamic structure of galaxy clusters, recovering key trends in pressure, temperature, electron density and entropy profiles across a broad range of cluster environments. While the agreement is particularly strong for dynamically evolved systems, the simulation also matches the profiles of disturbed clusters such as Coma and A119, where the large-scale ICM properties align well with observations.  

\item Current cosmological simulations, including SLOW, face difficulties in accurately reproducing core regions, particularly in electron density and entropy profiles, which tend to be systematically underestimated or overestimated. These discrepancies arise from a combination of numerical resolution constraints and limitations in the implementation of subgrid physics, including potentially excessive AGN feedback and incomplete modeling of non-thermal pressure support. The largest deviations are found in NCC clusters, where the spherical symmetry assumption used for profile de-projection is less applicable, and merger-driven turbulence and substructure interactions introduce additional complexities that are not fully captured. However, the ability to directly compare simulated and observed clusters on a one-to-one basis reveals a new pathway to identify where current models struggle to capture key physical processes, particularly in cluster cores, and provide critical insights into the mechanisms shaping core structure and the impact of mergers, turbulence, and feedback. Identifying these mismatches serves as a new tool to identify pathways toward refining AGN and star formation models with a stronger physical foundation.

\item The overall accurate reproduction of late-time profiles indicates that the simulation captures the formation histories of clusters well. This makes it possible to connect formation history to observables that are otherwise difficult to reproduce in simulations, such as the cores of galaxy clusters. We studied the evolutionary paths of the clusters of our simulation, classifying them based on their observed cores: CC, WCC, and NCC. We found that CC, WCC, and NCC clusters differ in their evolution, with CC clusters forming earlier and evolving more quiescently, while WCC clusters experience extended merger activity that disrupts their cores and increases entropy across their evolution, and NCC experience late-time major mergers that make them accrete most of their mass in later times. This link between formation history and core structure provides insight into regions of the ICM where subgrid models remain incomplete, particularly in the treatment of AGN feedback, turbulence, and radiative cooling.  

\end{enumerate}
These results underscore the value of constrained simulations in bridging theoretical models and observations, offering a powerful framework to investigate the formation and evolution of galaxy clusters in a cosmological context. By enabling direct comparisons between simulated and observed structures, we validate the ability of these simulations to recover the large-scale formation history of clusters while also exposing critical gaps in our understanding of cluster cores and their thermodynamic evolution. Our analysis highlights a clear connection between formation history and present-day core properties, with CC clusters forming earlier and evolving more quiescently, WCC clusters experiencing prolonged merger activity that gradually disrupts their cores, and NCC clusters undergoing late-time major mergers that significantly reshape their structure. This evolutionary link provides a new avenue for identifying the missing physical ingredients in subgrid models, particularly in AGN feedback, turbulence, and non-thermal pressure support. Moving forward, improving these models will be essential for capturing the full complexity of cluster cores and their thermodynamic states. Future high-resolution simulations, combined with increasingly precise observational data, will refine our understanding of these processes and further enhance the predictive power of constrained cosmological simulations, advancing our knowledge of galaxy cluster physics and large-scale structure formation in the Universe.




\begin{acknowledgements}

The authors thank the comments and discussions with all the members of this project. They thank the Center for Advanced Studies (CAS) of LMU Munich for hosting the collaborators of the LOCALIZATION project for a week-long workshop. This work was supported by the grant agreements ANR-21-CE31-0019 / 490702358 from the French Agence Nationale de la Recherche / DFG for the LOCALIZATION project. KD  acknowledges support by the Excellence Cluster ORIGINS, which is funded by the Deutsche Forschungsgemeinschaft (DFG, German Research Foundation) under Germany’s Excellence  Strategy – EXC-2094 – 390783311 and funding for the COMPLEX project from the European Research Council (ERC) under the European Union’s Horizon 2020 research and innovation program grant agreement ERC-2019-AdG 882679. The calculations for the hydro-dynamical simulations were carried out at the Leibniz Supercomputer Center (LRZ) under the project pn68na. We are especially grateful for the support of M. Petkova through the Computational Center for Particle and Astrophysics (C2PAP). 

\end{acknowledgements}

\bibliographystyle{aa}
\bibliography{example} %




\begin{appendix}

\section{Profile dependence on resolution}
\label{appendix1}

\begin{figure}[ht!]
\centering
    \includegraphics[width=0.9\linewidth]{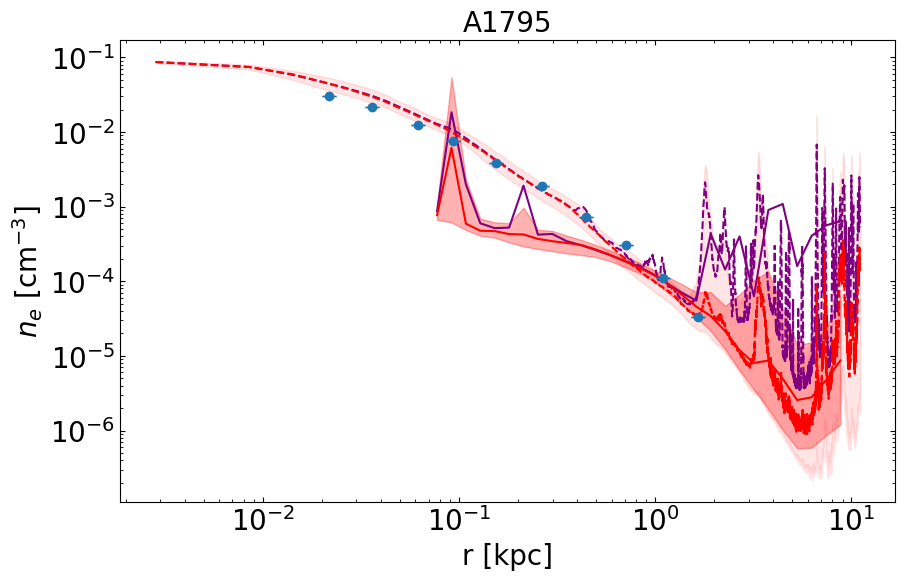}
    \caption{Normalized electron density profiles for the galaxy cluster A1795 at two different simulation resolutions. The red solid line and shaded region represent the median and $1\sigma$ scatter for the AGN1536$^3$ resolution run, while the red dashed line shows the median profile for the higher-resolution CR3072$^3$ run. The purple line represents the mean profile for AGN1536$^3$. Blue points correspond to observational data. The CR3072$^3$ simulation reproduces the electron density profile well across all radii, while the AGN1536$^3$ simulation underestimates the central density due to resolution limitations. The radius is normalized to the $R_{500}$  radius of the cluster.}
    \label{A1795 res}
\end{figure}

In this appendix, we examine the impact of simulation resolution on the electron density profiles of galaxy clusters. Figure \ref{A1795 res} shows the normalized electron density profiles of the simulated cluster A1795 at two different resolutions. The profiles are plotted as a function of radius normalized to the $R_{500}$ radius of the cluster.

The red solid line and shaded region correspond to the simulation run at AGN1536$^3$ resolution, while the dashed line represents the CR3072$^3$ resolution. The CR3072$^3$ simulation, which has higher mass and spatial resolution, successfully reproduces the electron density profile of the cluster, capturing both the core structure and the outer regions. In contrast, the AGN1536$^3$ simulation shows a clear deficit in the central density, failing to accurately resolve the dense core of the cluster. 

This discrepancy is likely largely driven by differences in resolution, which limit the ability of lower-resolution simulations to capture small-scale gas dynamics and central gas concentrations in cool-core clusters like A1795. However, we also acknowledge the presence of different subgrid physics between the two simulations: the CR3072$^3$ run is non-radiative and lacks cooling and feedback mechanisms, which can also influence the resulting profiles. Thus, both resolution and baryonic physics contribute to the differences observed.

The comparison highlights the necessity of high-resolution simulations when studying the inner regions of galaxy clusters, where the gas density gradients are steepest and accurate modeling is essential for reproducing observed thermodynamic profiles.

\end{appendix}
\end{document}